\documentclass[prb,preprint,floatfix]{revtex4-1} 

\usepackage{amsmath}  
\usepackage{amstext}
\usepackage{amsfonts} 
\usepackage{graphicx} 
\usepackage{hyperref}
\usepackage{cleveref}

\begin{document}


\title{Disruption of Supernovae and would-be ``Direct Collapsars''}

\author{John Middleditch}
\email{j.middleditch@gmail.com}
\affiliation{University of California, retired}

\date{\today}

\begin{abstract}

The speed of an \textit{intensity pattern} of polarization currents
on a circle, induced within a star by its rotating, 
magnetized core, will exceed the speed of light for a sufficiently 
large star, and/or rapid rotation, and will, in turn, generate focused 
electromagnetic beams which disrupt them. Upon core collapse within 
such a star, the emergence of these beams will concentrate near the 
two rotational poles, driving jets of matter into material previously 
ejected via the \textit{same} excitation mechanism acting through the 
pre-core-collapse rotation of its magnetized stellar core(s). This 
interpenetration of material, light-days in extent from the progenitor, 
produces a significant fraction of the total supernova luminosity, and 
the magnitude \textit{and time of maximum} of this contribution both 
vary with the progenitor's rotational orientation. The net effect is 
to render supernovae unusable as standard candles without further 
detailed understanding, leaving no firm basis, at this time, to favor 
\textit{any} cosmology, including those involving ``Dark Energy.''  
Thus we are not now, nor have we \textit{ever} been, in an era of 
precision cosmology, nor are we likely to be anytime soon. Mass loss 
induced through the same mechanism also keeps aggregates of gas and 
plasma in the early Universe, or at any other epoch, from forming the 
$\sim$billion solar mass stars which have been suggested to produce 
$\sim$billion solar mass black holes via ``direct collapse,'' but
can also provide a signature to predict core collapse some months
in advance.
We examine this mechanism through 
pulsar emission via polarization currents, in which the
emission power from any coaxial annulus of plasma decays only as 
1/distance for two exactly opposite rotational latitudes given by $\pm 
\arccos(\mathrm{c}/v)$, where c is the speed of light, and $v > 
\mathrm{c}$ is the speed of the rotating excitation.
We investigate  
why this effect results from circularly supraluminal excitations, 
as well as providing a discussion of, and further 
evidence for, the effect in the Parkes Multibeam Pulsar Survey.  

\end{abstract}

\maketitle 

\section{Introduction} 
\label{sec:intro}

An obliquely magnetized neutron star with an angular velocity of $\omega$ 
radians/s, will excite polarization currents in the surrounding plasma at 
projected radius, $R$, with a circumferential pattern, which, for $\omega 
R > \mathrm{c}$, exceeds the speed of light. The
model\cite{Ar94,H98,AR03,AR04,AR07,AR08,AR08a} of pulsar emission, which 
takes such supraluminally excited polarization currents into account, is 
the only model thus far to predict the observed emission bands and their 
increasing spacing with radio frequency\cite{HE07} in the GHz spectrum of 
the Crab pulsar interpulse, in addition to matching its observed spectrum 
out to the $\gamma$-ray — over 15 orders of magnitude (see also [9]).

\begin{figure}[ht!]
\centering
\includegraphics[height=11cm,width=14.1856cm]{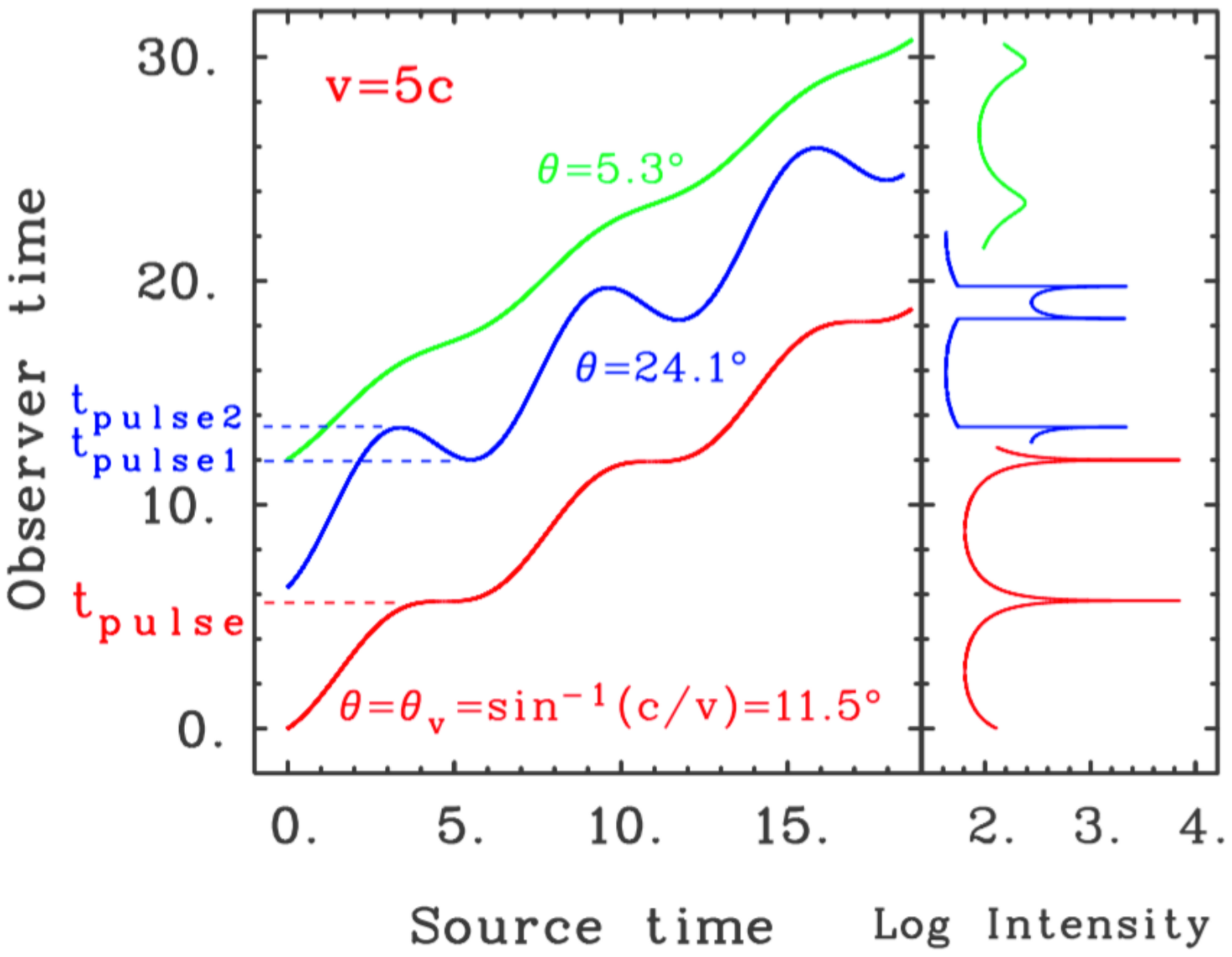}
\caption{
From [\onlinecite{M12}]. Left: Observer time as a function of Source time 
for a supraluminally induced circular excitation at 5 times light speed, 
with a frequency of $\omega = 1$ radians/s, or a period of $2\pi/\omega = 
2\pi$ seconds, at a radius $R = 5 \mathrm{c}/\omega$, where c is the speed 
of light, for three spin co-latitudes, $\theta$ (polar angles).  The three curves have been arbitrarily offset by a few units for clarity.
The difference between times for an observer on the \textit{distant} 
($+Y,Z$) plane and the source on the R-circle (centered at the origin, 
$(X,Y)=(0,0)$) can be expressed as: $t_{\mathrm{obs}} = t_{\mathrm{src}} - 
R/\mathrm{c}~\sin(\omega t_{\mathrm{src}})\sin(\theta)$.
Its derivative with respect to time is given by:
$\partial t_{\mathrm{obs}}/\partial t_{\mathrm{src}} = 1 - 
(\omega R/\mathrm{c})~\cos(\omega t_{\mathrm{src}}) \sin(\theta)$.
The second derivative,
$\partial ^2 t_{\mathrm{obs}}/\partial ^2 t_{\mathrm{src}} = 
(\omega ^2 R/\mathrm{c})~\sin(\omega t_{\mathrm{src}}) \sin{\theta}$,
is zero twice every period, and every other zero is a point of inflection, 
provided $\theta = \arcsin(\mathrm{c}/(\omega R))$, making the first
derivative 0. The other zeros mark the times of maximum slope.
Right:  the integral, for 65,536 Source times per cycle, into
512 discrete bins per cycle of Observer time/phase, for the three curves 
(i.e., pulse profiles).
}
\label{rounda}
\end{figure}

\begin{figure}[ht!]
\centering
\includegraphics[height=13.3798cm,width=17.2694cm]{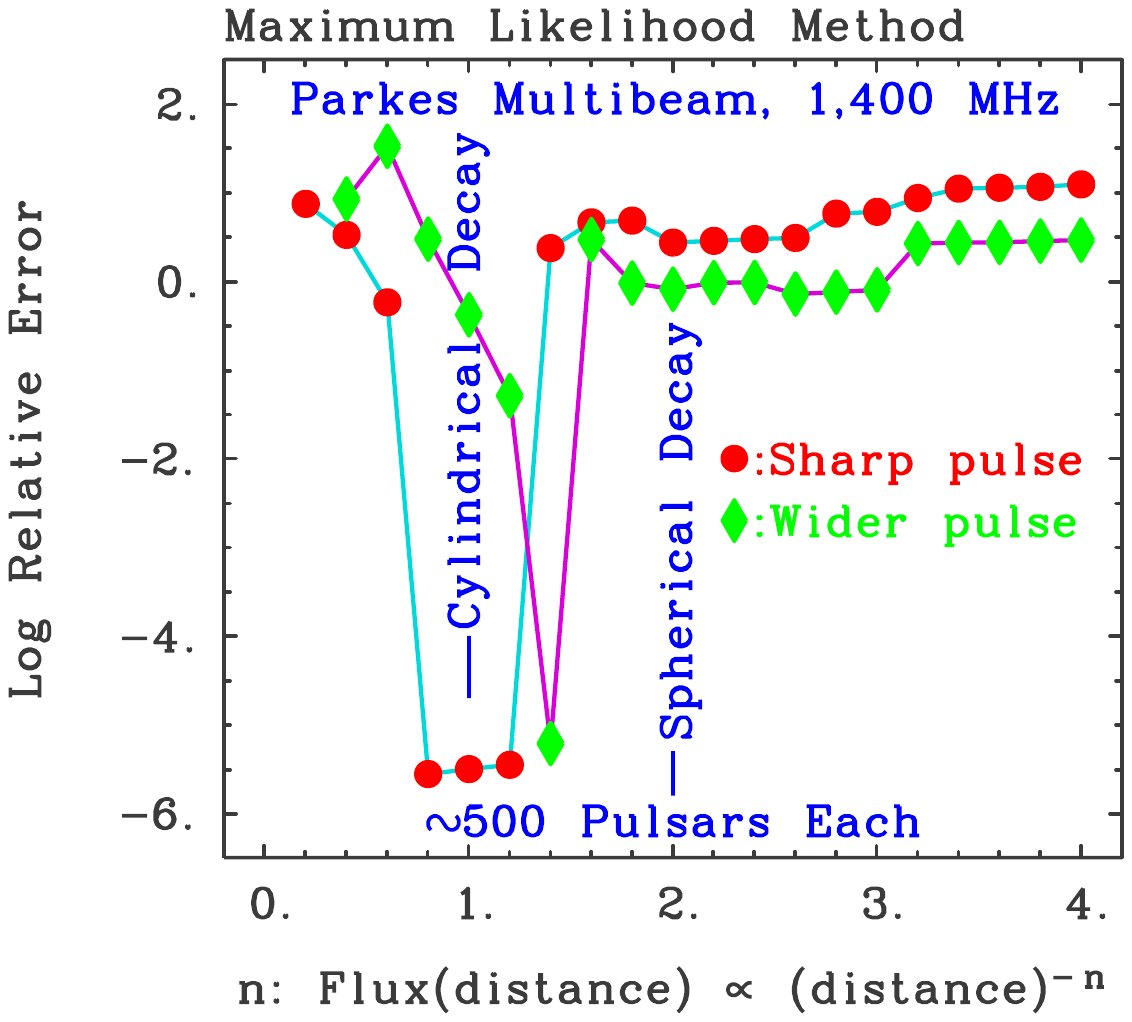}
\caption{The Maximum Likelihood Method of [\onlinecite{EEP}] as applied
in [\onlinecite{MLM9}] to the Parkes Multibeam Survey pulsars without any
filtering, and then to the half of that sample whose pulse widths are less 
than 3\% of their pulse periods.
}
\label{dlaw}
\end{figure}

This model also predicts that the emitted power will obey a 1/distance 
law for two opposite spin latitudes on the sky, which approach, at large 
distances, $\pm \arccos(\mathrm{c}/v)$, where c is the speed of light, 
and $v=\omega R >\mathrm{c}$ is the velocity of the update of the
circumferential polarization current pattern. At these latitudes, the 
curve of the Observer time as a function of Source time is cubic, i.e., 
there is a point of inflection for one (periodic) value of Observer 
time, where extended ranges of Source times map onto any small interval
centered at these Observer times (Fig.~2 of [\onlinecite{H98}]).  As a 
result, a single, very sharp peak appears\cite{two} in the observer's 
pulse profile when near that time/phase (see the lowermost curve in 
Fig.~\ref{rounda}). Thus, for any given pulsar,\cite{wind} observers 
can tell when they are at, or near, one of these two favored latitudes 
by the pulse profiles that they record. At more equatorial latitudes,
the single pulse splits into two separate pulses, while at more polar 
latitudes, the single pulse broadens and weakens (the top curve of 
Fig.~1). See also Fig.~2.10 of [\onlinecite{ACS}].

\begin{figure}[ht!]
\centering
\includegraphics[height=15cm,width=19.3440cm, trim={1cm 2cm 0cm 
1.1cm}]{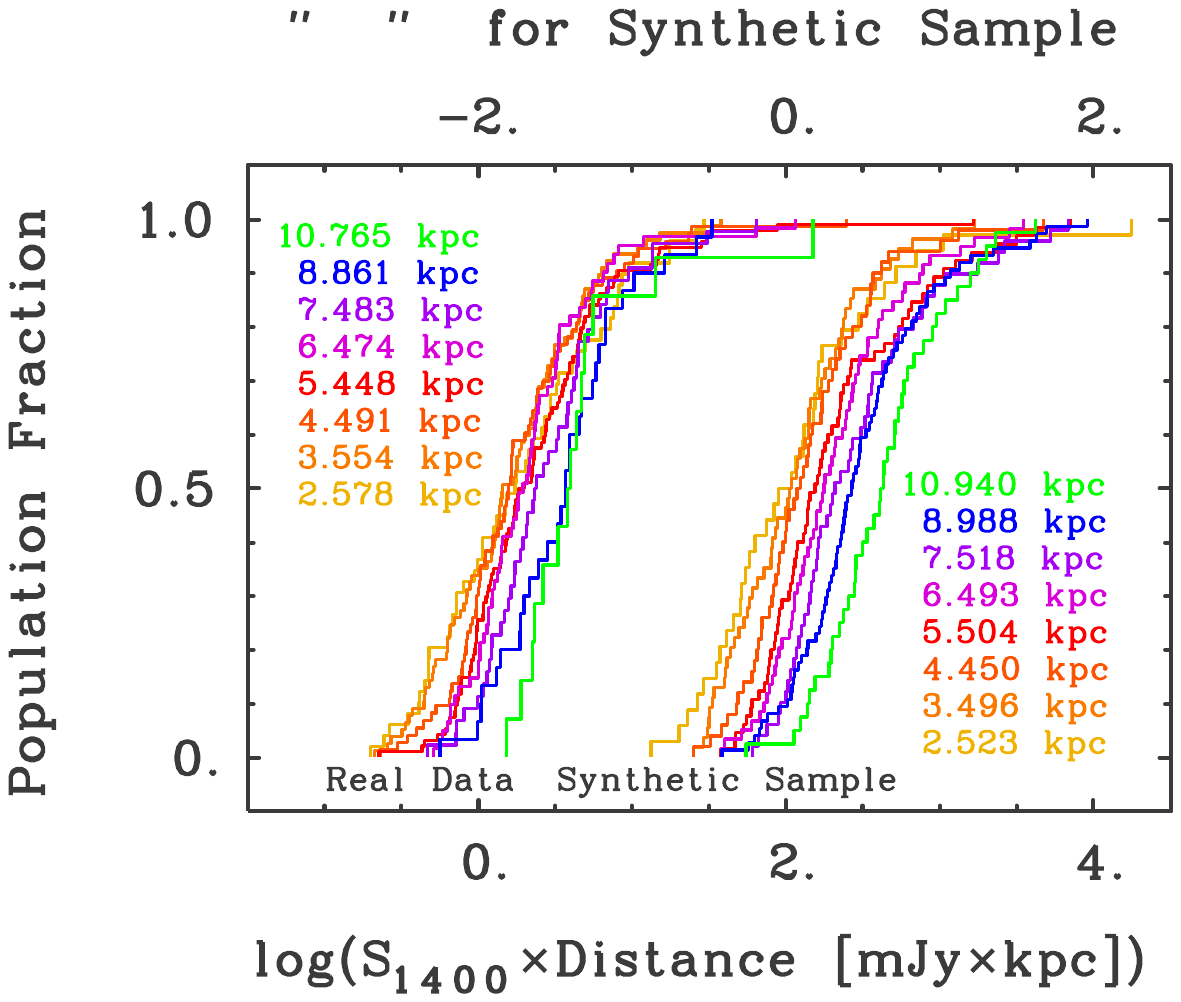}
\caption{
From [\onlinecite{M12}]. Left, lower horizontal axis:  the cumulative 
populations of 497 pulsars from the Parkes Multibeam Survey with pulse 
widths less than 3\% of their periods, which fall into binned distances 
with centers between $\sim$2 and $\sim$11 kpc, plotted against the log 
of the product of their 1400 MHz fluxes and their NE2001 distances from 
[\onlinecite{Cordes02}] and [\onlinecite{Cordes03}]. The spread of the 
different flux-times-distance products at a given population fraction 
diminishes as their components become more luminous (and less detection-
limited), indicating a 1/distance law. Right, upper horizontal axis:  
the same for a synthetic population of pulsars whose fluxes obey the 
inverse square law, and the spread in the flux-times-distance products 
does not diminish (but would \textit{collapse} for flux-times-distance-
squared products). The two families of curves are both in order of 
increasing distance from left to right at a population fraction of 0.2.
}
\label{oneovrppb4}
\end{figure}

In the section that follows immediately (Section \ref{Old and New}), 
new evidence is added to the old for the effects of supraluminally 
induced polarization currents in pulsars. Section \ref{violate} 
derives the 3-dimensional path of the focused beams produced by 
circularly supraluminal excitations, and discusses their mathematics.  
Section \ref{Further} continues with a few more mathematical notes.  
A discussion of polarization effects follows in Section \ref{Pol}.

Section \ref{MI} discusses the implications for known pulsars, and
then describes how annuli of induced polarization currents 
throughout a large star produce focused beams which break the 
stellar surface very close to its rotational poles. This
section then describes the implications, drawn in [\onlinecite{M12}] 
from early photometric data of SN 1987A, about the interpenetration
of the near polar focused beams through circumstellar material.

Section \ref{AGN} discusses distance effects in pulsars, GRBs, 
AGN jets, and binary few-million solar mass black holes.
Section \ref{HTRO} discusses GRBs as a result of other kinds of
supraluminal excitations, deriving redshifts from their 
afterglows, related effects of NS-NS and BH-NS mergers, and the 
possibility of focused \textit{gravitational}
beams. Section \ref{Pulsars} gives a general discussion of 
supraluminal excitations and focused beams and their effects
on pulsars, supernovae, the Sun, Hercules X-1, and other objects.
Section \ref{Conc} concludes.

\section{Old and New Evidence}
\label{Old and New}

Evidence for this effect in the Parkes Multibeam Pulsar 
Survey\cite{pksmb} was first presented in 2009, but a more recent 
attempt by another group\cite{SD2016} to duplicate the results, 
based \textit{roughly} on the same Maximum Likelihood Method (MLM -- 
[\onlinecite{EEP}]) employed in 2009, failed to find the effect.  
The ``Stepwise'' MLM does not rely on a simple functional form for 
the luminosity function, $\phi$(L), but instead iteratively 
determines $\phi_\mathrm{k}$ for the k$^\mathrm{th}$ equal step in 
luminosity. It defines the logarithm of the likelihood function by 
the logarithm of the sum (over all sources) of the logarithms of the 
$\phi_\mathrm{k}$ every time a source falls into the k$^\mathrm{th}$ 
luminosity bin, \textit{minus}, the sum (over all sources) of 
logarithms of, a sum over all $\phi_\mathrm{j}$ for all the 
j$^\mathrm{th}$ 
luminosity bins above the minimum luminosity defined by an 
\textit{unspecified} function of the redshift of the source. The 
inner kernel of this subtracted term is also multiplied by the 
width of all the steps in luminosity, $\Delta$L, for dimensional 
consistency. A constraint is also added with a Lagrange multiplier. 

Needless to say, the adaptation of such a method, more appropriate 
to clusters of sources, than to pulsars within the Milky Way disk, 
is non-trivial. This more recent study of [\onlinecite{SD2016}] 
included only the $\sim$half of those pulsars in the sample with 
pulse profiles whose peak widths were \textit{greater} than 3\% of 
their pulse periods,\cite{3} unlike those in the bottom curve in 
Fig.~1, and in doing so used pulsars for which there was no 
expectation of a 1/distance law. The 2009 study only excluded 
these after submission. The results for this restricted sample of 
497 pulsars 
were much more dramatic, as can be seen in Fig.~\ref{dlaw}. 
A good guess as to the source of the disagreement is the 
difference in the number of bins chosen, with the smaller number 
being more desirable to show the effect in a limited data set, as 
can be seen in Fig.~\ref{oneovrppb4}, where cumulative populations 
were made for eight different subsets over a factor of four in
distance. Meanwhile, it is far more important to establish the 
validity of the violation of the inverse square law independently 
of the binning issues of the MLM. If need be, the sets from 
Fig.~\ref{oneovrppb4} used in the MLM method will likely
reproduce the result of Fig.~\ref{dlaw}.

\begin{figure}[tbp]
\centering
\includegraphics[height=14.4762cm,width=18.0544cm, 
trim={1cm 2cm 1cm 1.1cm}]{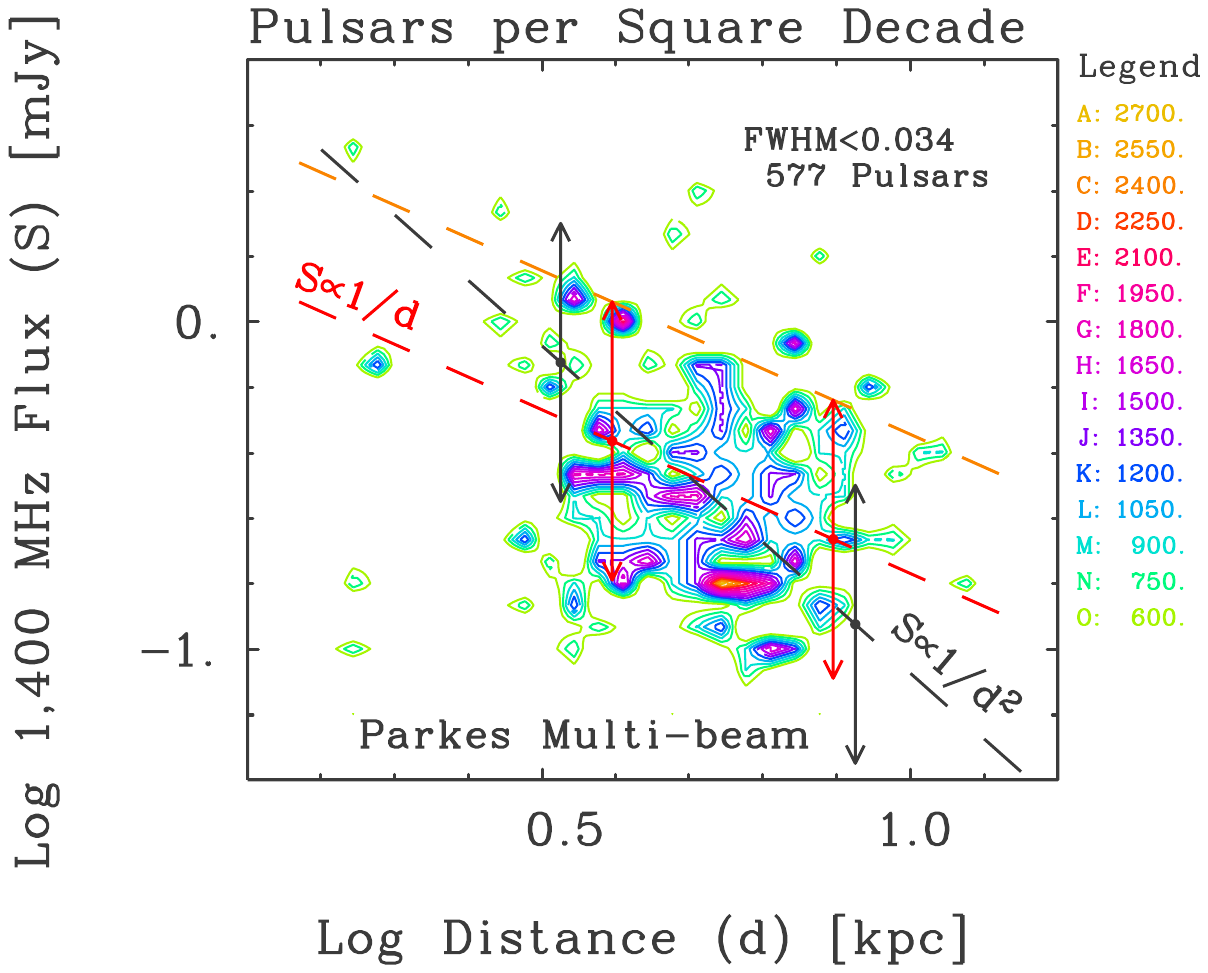}
\caption{
A contour plot of 577 pulsars from the Parkes Multibeam Pulsar Survey 
whose pulse profiles consist of a single sharp peak with a half width 
in time less than 3.4\% of the period ($\mathrm{F0} \times \mathrm{W50} 
< 34$). The numerical slope of the two parallel dashed lines (red and 
orange) is -1, corresponding to a 1/distance law, while that of the 
steeper dashed line (black) is -2, corresponding to an inverse square 
law. The arrows show what the span of the luminosity function would 
be if the opposite pairs were appropriately centered (which the red 
arrow pairs are -- slope of -1 -- and the black pairs are not -- 
slope of -2).
}
\label{ppb}
\end{figure}

It is actually easy to show that the narrow-pulse-profile-restricted 
sample trends toward a 1/distance law as the pulsar fluxes times 
distances increase above the detection-limited values below 2 mJy-kpc 
(see Fig.~\ref{oneovrppb4}). While the mJy-kpc spread of the curves 
of the actual data sample (at the left) trends toward zero at high 
mJy-kpc/population-fraction, a signature of a 1/distance law, the 
synthetic sample (at right), formulated from an inverse square law, 
retains its spread of $\sim$0.6 (a factor of 4 between 2.5 and 
$\sim$10 kpc) since it would require multiplication by another 
factor of distance to collapse.

\begin{figure}[tbp]
\centering
\includegraphics[height=15cm,width=19.344cm]{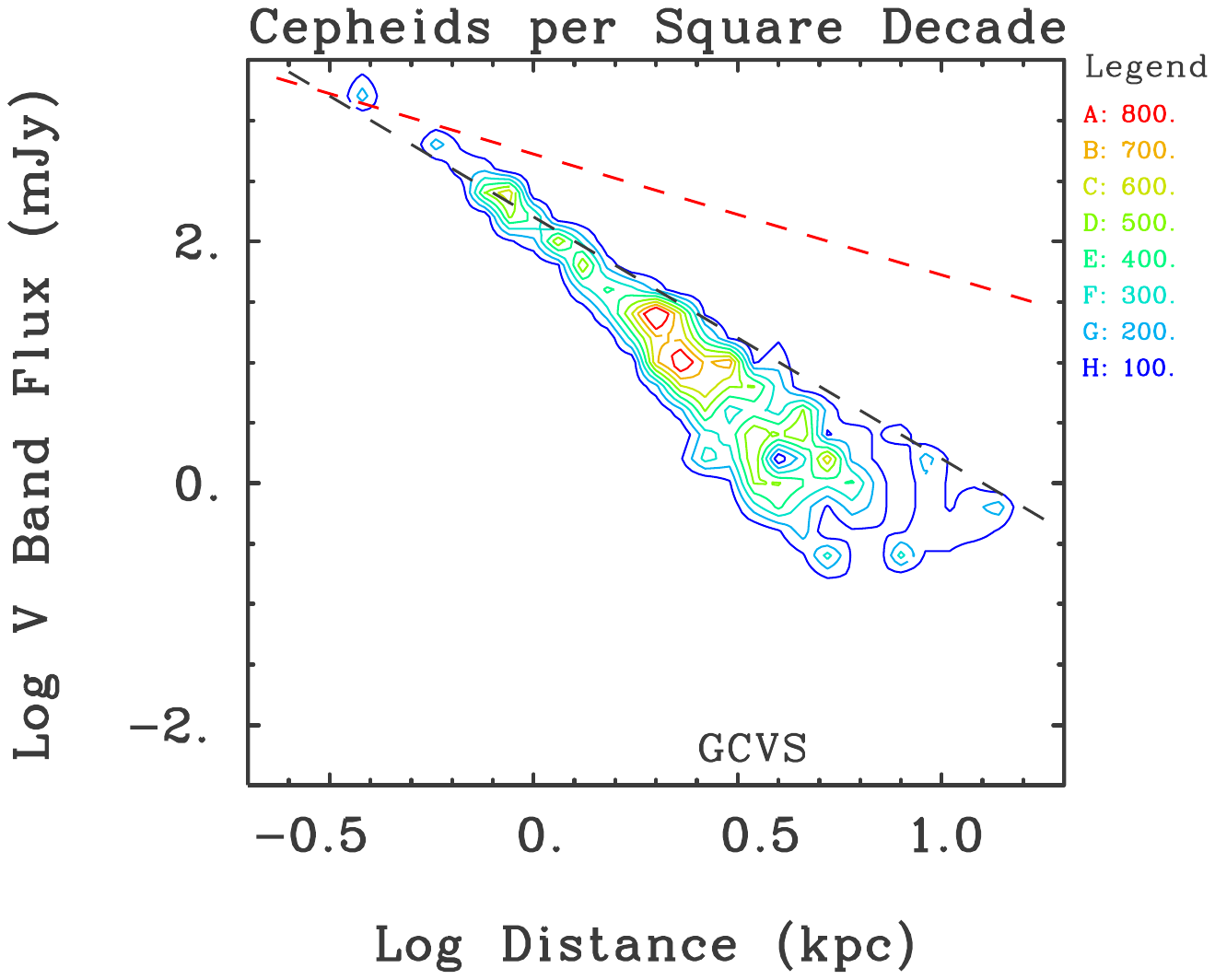}
\caption{
A similar contour plot of 407 Cepheid variables. The numerical 
slope of the (mostly upper) dashed line (red) is -1, corresponding 
to a 1/distance law, while that of the steeper line (black) is -2, 
corresponding to an inverse square law. The Cepheids actually 
track slightly steeper with distance than a power law of -2, a 
sign of interstellar extinction.}
\label{ceph}
\end{figure}

The \textit{population} is dense enough (Fig.~\ref{ppb}) so that 
its upper cutoff (orange dashed line) on a log flux-log distance 
graph follows a numerical slope of -1 well, and a slope of -2 poorly, 
lying below/above the steeper (black) line at small/large distances.  
The black arrows show that the inverse square law line does not cut 
through the same fraction of the luminosity function in the way that 
the 1/distance law (red arrows) does well. This conclusion is 
susceptible to having larger populations near one of the two extremes 
of distance as opposed to mid-distances, but this is clearly not the 
case, however, as the densest part(s) of the distribution are located 
at mid-distances. Indeed, it is remarkable how well this 
distribution's upper boundary follows a 1/distance law, in spite of 
it not having been corrected for its denser population at 
mid-distances. By its very nature, Fig.~\ref{oneovrppb4} does not 
suffer from this problem.

We can check this contour-plotting method on other objects which are 
known to follow an inverse square law, by analyzing its response to 
the Cepheids in our Galaxy. Named after the variable star, $\delta$ 
Cephei (the fourth brightest star in the constellation, Cepheus -- 
``the King''), the Cepheids are massive stars whose period of 
variability has a near linear relationship to their luminosity. Thus, 
once we know the interstellar extinction along the line of sight to 
the particular Cepheid (see, e.g., [\onlinecite{amores}]), we can 
infer its distance from the difference between its absolute and 
apparent magnitudes. Figure \ref{ceph} confirms the contour plot 
reflects an inverse square law for the Cepheids.

\begin{figure}[ht!]
\centering
\includegraphics[height=14cm,width=18.0544cm]{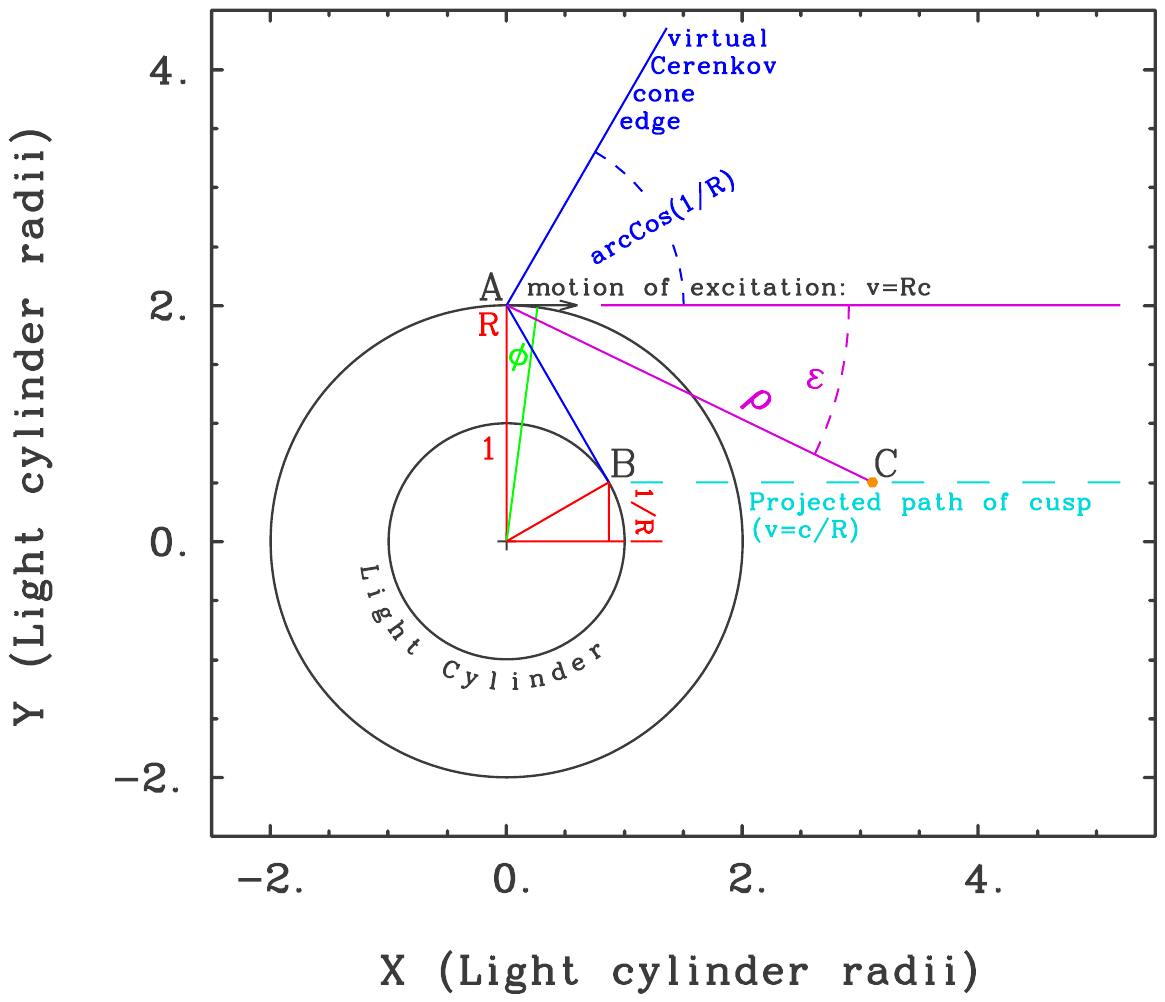}
\caption{
The geometry of circularly supraluminal motion. The instantaneous 
velocity of $R$ times the wave speed (of light or some other 
propagating wave) is in the +X direction at point `A,' $(X,Y,Z)$ = 
(0,$R$,0). The variable, $\phi$ measures the clockwise angle along 
the circle of radius $R$. The virtual Cerenkov cone, whose axis is 
parallel to the X axis, but displaced in Y by $+R$, is viewed from 
above the X-Y plane of the graph, so that its edges form the upper 
line from point `A', and the lower line from `A' to `B,' on the 
circle where the rotation of angle due to the motion on the
$R$-circle is exactly the wave speed (i.e., the ``Light Cylinder'' 
when extended normal to the X-Y plane). The dashed line at $Y =1/R$ 
is the projection onto the X-Y plane of the (hyperbolic) path of the 
cusp generated by the supraluminally updated polarization currents 
near point `A'.
}
\label{cuspg}
\end{figure}

\section{How and why the inverse square law is violated}
\label{violate}

Supraluminal excitations may result in a focusing of an increasing, 
extended interval of Source time onto a small interval of Observer 
time, provided the excitation is accelerated.\cite{Lilly} Any 
circular excitation of constant speed is always accelerated toward 
the center of its circle (see Fig.~\ref{cuspg}). If there is to be 
any focusing of such a supraluminal excitation (a point source 
polarization current traveling in a circle), the focus must lie on 
the cone whose apex coincides with the source, with its axis, and 
opening, colinear with the instantaneous direction of motion, and 
whose half angle is given by $\arccos(\mathrm{c}/v)$, where c is the 
speed of light (in whatever medium) and $v$ is the speed of the 
excitation. This opening angle is the compliment 
($\arcsin(\mathrm{c}/v)$) of that of the usual Cerenkov cone. For 
convenience, we will label this as the ``virtual Cerenkov cone,'' or 
``vCc."

\begin{figure}[ht!]
\centering
\includegraphics[height=16.5cm,width=21.2784cm]{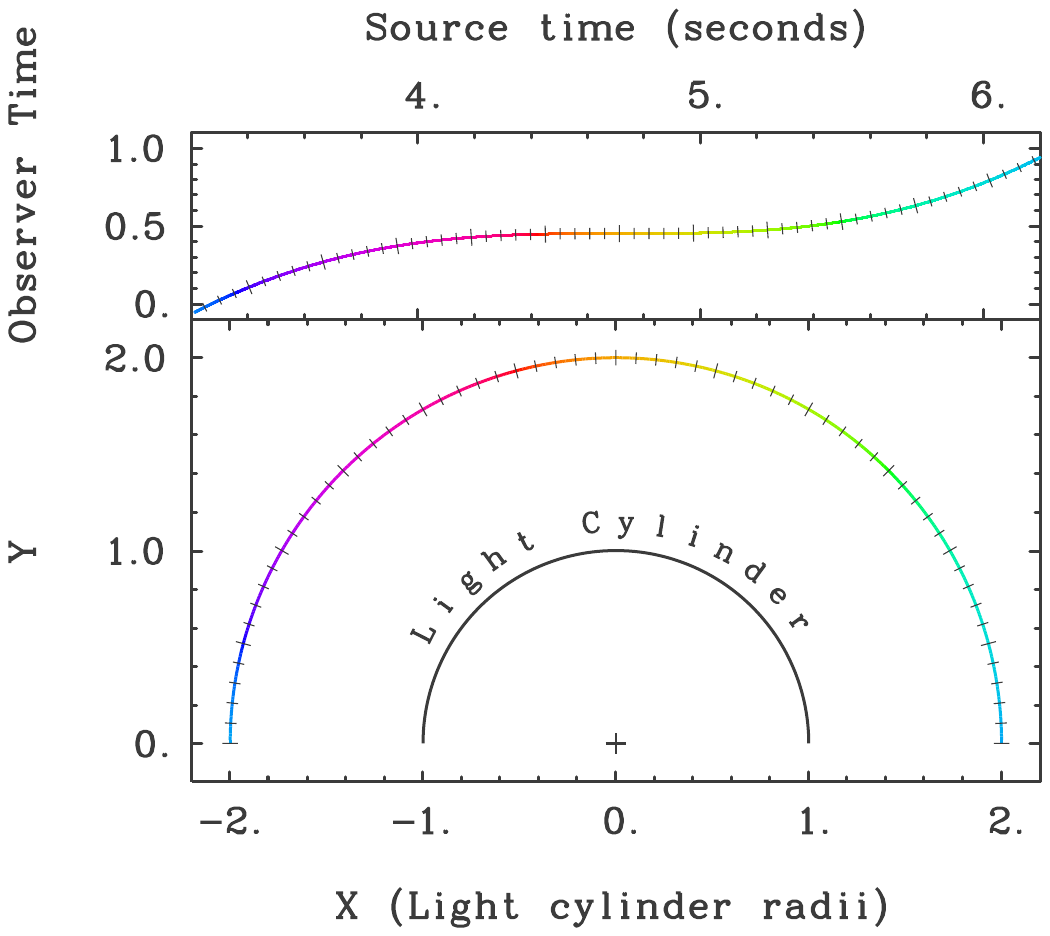}
\caption{(Lower frame) As in Fig.~\ref{cuspg}, the geometry of 
circularly supraluminal motion. The instantaneous velocity is 
clockwise -- in the +X direction at the top of the outer (colored) 
circle with minor/major tick marks every 3/15 degrees. (Upper frame) 
The color-coded contributions, of polarization currents in the outer 
circle/annulus of the lower frame, for a pulsar with a period of 
$2\pi$, to the Observer time vs.~Source time curve (which is also 
plotted in Fig.~\ref{roundopsi}, in green).
}
\label{xtry}
\end{figure}

Any temporal information in a part of the source that travels toward 
the observer at the speed of light and with zero acceleration 
collapses onto a single arrival time, as can be seen in 
Fig.~\ref{xtry}. The \textit{projected} separation of the sources 
along the line of excitation is preserved in the outgoing wave, but 
the lessening distance penalty for perpendicular offset reduces the 
difference in their arrival times at increasingly greater distance. 
The same mechanism for acoustic waves is responsible for the ``Sonic 
Boom.'' Centripetal acceleration of the sources toward the center of 
a circle will always, for the right radius, do better than the Sonic 
Boom, as we will see below. On the other hand, if the observer is not 
on the vCc of the excitation, the effects of many nearby sources on 
the path of excitation will arrive at different times, no matter what 
the distance.

The virtual Cerenkov cone of circular motion at $R$ times the wave 
speed in the medium will always have a point tangent to the smaller, 
concentric circle with a radius, $R_\mathrm{LC}$, where the angular 
motion of $\omega$ radians per second corresponds to exactly the wave 
speed, or $\omega R_\mathrm{LC} = \mathrm{c}$ for electromagnetic 
radiation. Extension of this circle in the directions of its normals 
becomes the ``Light Cylinder." For convenience we will use dimensions 
so that $R_\mathrm{LC} = 1$. The actual units may be re-established 
by multiplying the constants, including functions of $R$, by powers of 
$R_\mathrm{LC}$ to make their units consistent, within each equation, 
with those of the powers of the variables, $X$, $Y$, and/or $Z$, each 
of which has units of length.

\begin{figure}[ht!]
\centering
\includegraphics[height=16.5cm,width=21.2784cm]{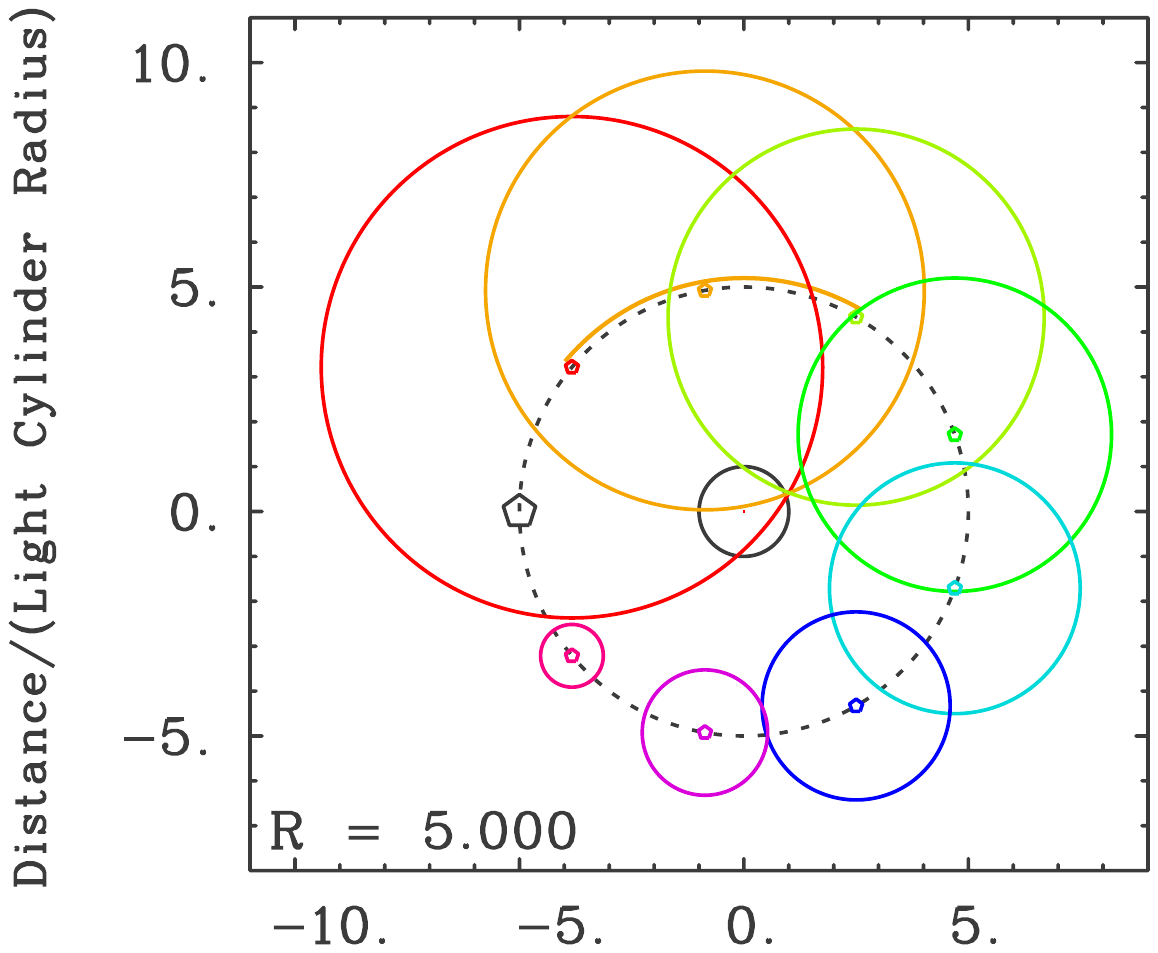}
\caption{A series of nine Huygen's wavelets produced by an excitation 
ending at the left part of a circular path (pentagon) having moved 
clockwise at five times the wave speed around the circle. The sources 
from 10 h 20 m to 1 o'clock (an 80$^{\circ}$ span -- highlighted by 
the orange arc) all contribute to the focus just past 2 o'clock on 
the circle of radius 1.
}
\label{superl}
\end{figure}

With the help of Fig.~\ref{superl}, which has a tangent point on the 
$R~=~1$ circle at ~~21.$^{\circ}$5 ($\arccos(1/5) = 11.^{\circ}537, + 
10^{\circ}$ from the point at 11 hr 40 min being 20 min short of 12 
hr) we can see that the tangent point 'B' in Fig.~\ref{cuspg} 
($30^{\circ}$ for $R = 2$ on the small circle) is likely the first 
point of the focus curve we are seeking. If we start with a point `A' 
on the X-Y plane at (0,$R$) in Fig.~\ref{cuspg}, where the 
supraluminal excitation is traveling at $R$c in the +X direction 
(i.e., rotating ``clockwise''), the vCc appears in profile, with its 
axis parallel to the X axis. This cone is tangent to the light 
cylinder at point `B': 
\begin{equation}
\label{pointBxyz}
(X,Y,Z) = (\sqrt{1 - {1 \over R^2} },{1 \over R},0), 
\end{equation}
and in consequence the line from `A' to `B' coincides with the lower 
edge of the vCc. The equation for this line is:
\begin{equation}
\label{linevCc}
Y = -(\sqrt{R^2 - 1}) X + R,
\end{equation}
and we will need it to relate an arbitrary point at ($X$,$Y$), to the 
right of point `B' in Fig.~\ref{cuspg}, to the hyperbola on the vCc 
at the given $Y$ value, in order to solve for its $\pm Z$ values. 
Given the $Y$ value, the $X$ 
solution to Eq.~\ref{linevCc}, $X_o$, is the $X$ fiducial in the
equation for the corresponding hyperbola on the vCc which returns the 
value for $Z$:
\begin{equation}
\label{Z(X,X_o)}
Z = \pm \sqrt{R^2 -1} \sqrt{X^2 - {X_o}^2}~;~X_o = (R - Y)/\sqrt{R^2-1} 
~.
\end{equation}

In order to find the exact path which will mark the location of the 
cusp/focus on the vCc (or confirm that it is an hyperbola), we must 
compute the distance, $\rho$, from a point on the $R$-circle near `A' 
to the cusp/focus point on the vCc, as a function of $X$, and $Y$, for 
a range of angles, $\phi$ (Fig.~\ref{cuspg}), on the $R$-circle, and 
test for a  $\rho(\phi)$ function that changes by only a small 
fraction of a light-radian over a large range in $\phi$ (of up to a 
radian), which is exactly what we need to get many Source times mapped 
into one, or a limited range of, Observer time. Thus, starting at an 
arbitrary point on the $R$-circle near `A', at phase $\phi$, or 
$(R\sin{\phi}, R\cos{\phi}, 0)$, and computing the distance to near 
point `B', or ($\sqrt{1 - 1/R^2}$,$1/R$,0), we get:
\begin{equation}
\label{ro(RXY)}
\rho(X,Y,\phi) = ((RX)^2 - 2RX \sin{\phi} + 2RY (1 - \cos{\phi}) )^{1/2}.
\end{equation}
We notice that terms in $Y^2$ and others that do not multiply 
trigonometric functions of $\phi$ have vanished, leaving a simple $Y$ 
dependence for the factor, $(1 - \cos{\phi})$.

By removing a factor of $RX$ from the square root and after using the binomial 
expansion for the 1/2 power, we get:
\begin{equation}
\label{ro2(RXY)}
\rho \sim RX~( 1 - ({{\sin{\phi}} \over {RX}}) + {{RY} \over (RX)^2} (1 - \cos{\phi})
- {1 \over 2} (\cdots)^2 + \mathrm{higher~terms} ),
\end{equation}
where the `$\cdots$' represents the grouped terms following the initial `1'. 
A factor of 1/4 from the binomial expansion coefficient has been absorbed by 
the choice of `$\cdots$' when squared. The first term, $RX$ after 
multiplying through the parenthesis to the `1', represents the macroscopic 
distance to the location of the beam focus.  

The -$\sin{\phi}$ in the next 
term expands (in radians) initially to $-\phi$, i.e., the distance, $\rho$, 
is less for higher positive $\phi$ values, which advance any excitation
near point `A' in Fig.~\ref{cuspg} to higher $X$ toward the right and the 
observer/beam focus. This `$-\phi$' cancels the delay of the 
source rotation on the $R$-circle, on which motion toward positive $\phi$ 
costs time, allowing many source points on the $R$-circle to nearly 
simultaneously affect one observer point. However, since Eq.~\ref{ro2(RXY)}
is an approximation, an error in this linear term in $\phi$ with a magnitude
which is inversely proportional to distance is always present, and in 
consequence prevents any signal from becoming infinite while producing the 
1/distance law. 

The next term in the expansion of the $-\sin$ is ${\phi}^3 / 6$. This 
cubic term has been mentioned above, and its slow departure from 0 near 
$\phi = 0$ would lead to an infinite response at one particular Observer 
time for every cycle, if it were not for the residual linear term in 
$\phi$.

We can see how the infinity would be generated by estimating the response 
at the observer position, which, is how much Source time (essentially 
$\phi$, in radians) gets mapped into a small interval of Observer time.  
In effect, the response is the derivative of Source time as a function of 
Observer time.  If high, then a large interval of Source time maps onto a 
small interval of Observer time.  Our $f(\phi) = {\phi}^3/6$, as shown by 
the bottom curve in the left hand frame of Fig.~\ref{rounda}, however, is 
the Observer time as a function of Source time  -- the \textit{inverse} 
of the function we need. We must turn the function on its side, as 
Fig.~\ref{rounda} suggests. Thus we need the derivative of $\phi(f)$:
\begin{equation}
\label{phi(f)}
\phi(f) = (6f)^{1/3}.
\end{equation}
Differentiating we get:
\begin{equation}
\label{d(phi)/df}
{{\partial \phi(f)} \over {\partial f}} = 2^{1/3}(3f)^{-{2/3}},
\end{equation}
which is infinite for $f=\phi=0$.

Continuing, there are more terms from the `$1 - \cos(\phi)$', 
of which the lowest one is, when multiplied by the leading $RX$,
\begin{equation}
\label{phi^2RYterm}
{RY \over 2{RX}}{\phi}^2,
\end{equation}
and a similar term, 
\begin{equation}
\label{phi^2RXterm}
{-{\phi}^2 \over 2{RX}}, 
\end{equation}
comes from the `$\phi$' part of the `$\sin$' term in the `$-{1 \over 2} 
(\cdots)^2$' continuation of the binomial expansion. If $Y=1/R$, the same 
value as the $Y$ for point `B', these terms cancel completely, and 
permanently if $Y$ stays at $1/R$, as do the factors which do not 
multiply functions of $\phi$.

Thus the focus, which starts at $Y=1/R$ at point `B', is always at $Y=1/R$.  
The next higher term in $\phi$ also comes from two contributions, and for 
$Y=1/R$, after multiplying through by the leading $RX$, amounts to 
\begin{equation}
\label{phi^4term}
{{\phi}^4 \over {8RX}}, 
\end{equation}
which, because it is a higher power than the cubic, \textit{and} 
decreases with distance, is not important.

\begin{figure}[ht!]
\centering
\includegraphics[height=15.0cm,width=19.344
cm]{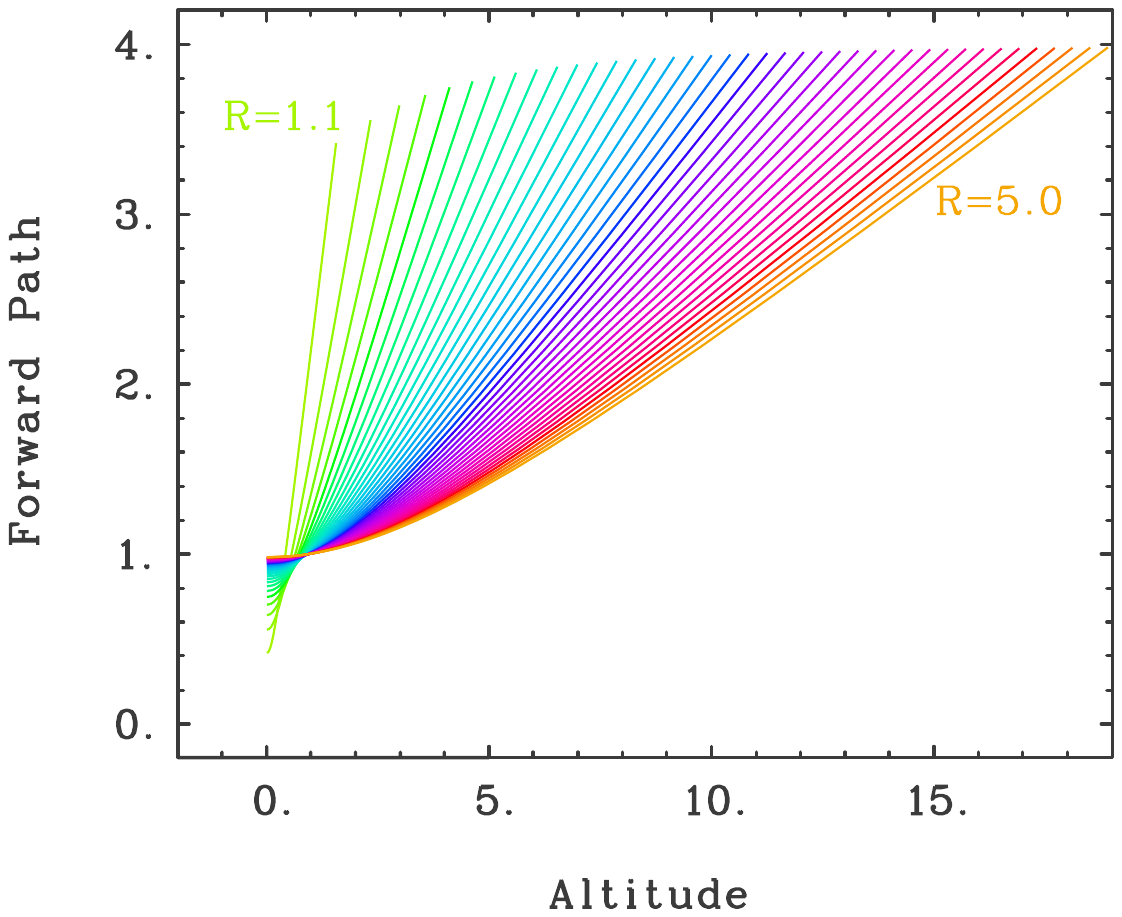}
\caption{
Curves of the path of the cusp plotted in the Z-X plane for 
$R = 1.1~\mathrm{to}~5.0$ in steps of 0.1.
}
\label{ZX}
\end{figure}

Since the Y value for the path of the focused beam remains at $1/R$ out to 
infinity, the equation for the focused beam on the vCc is an hyperbola 
which is simple to express:
\begin{equation}
\label{Z(XR);Y(R)}
Z =\pm  \sqrt{R^2 - 1} \sqrt{X^2 - 1 + {1 \over R^2} }; ~Y = {1 \over R} .
\end{equation}
Figure \ref{ZX} plots these curves in Z-X for $R = 1.1~\mathrm{to}~5.0$.

\begin{figure}[ht!] 
\centering
\includegraphics[height=15cm,width=19.344cm]{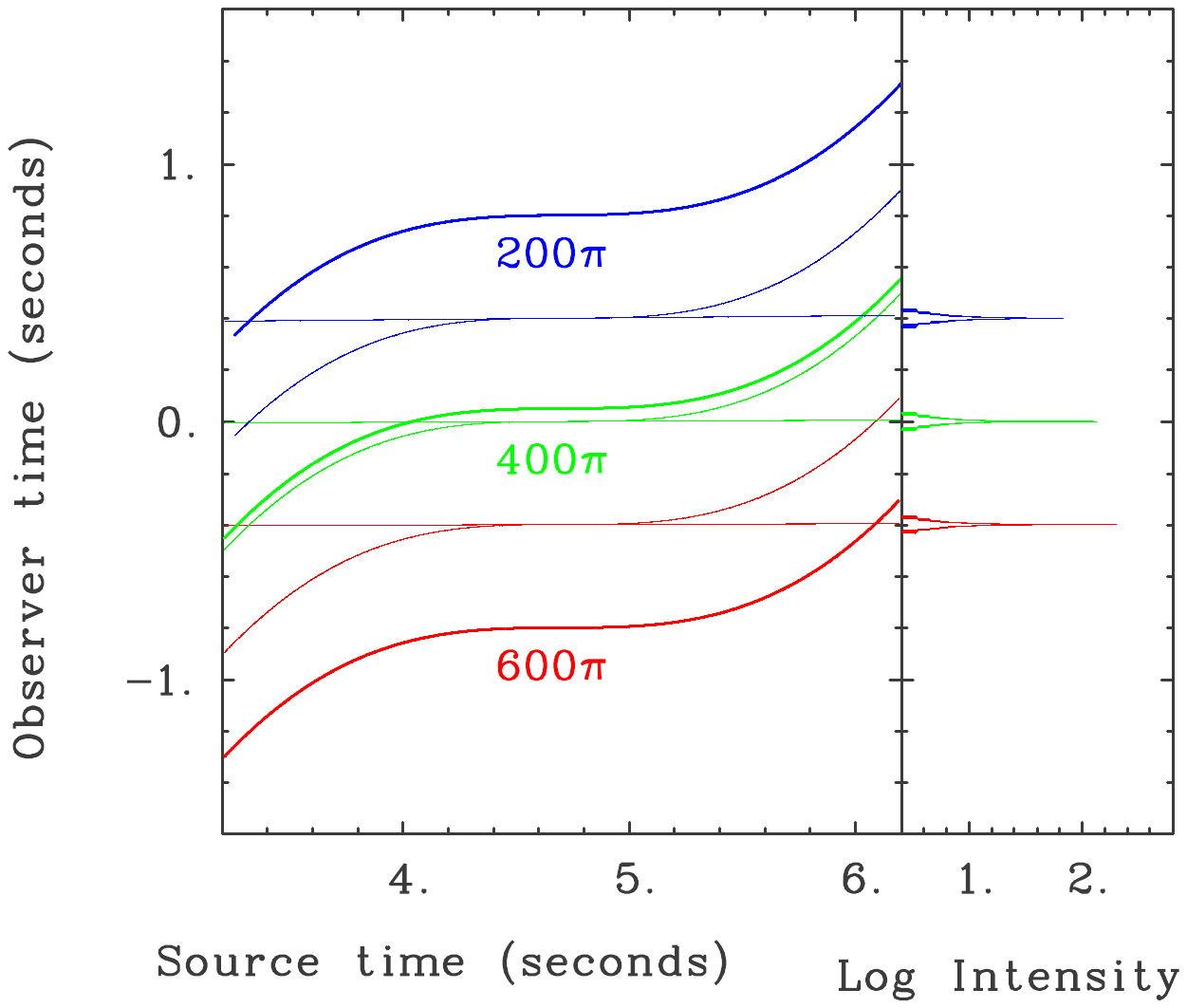}
\caption{Observer time as a function of Source time curves for
a \textit{simulation} of a pulsar with a period of 2$\pi$ and plasma 
at $R=5$, for three different distances, $200\pi$ (top two [blue]
curves), $400\pi$, (middle [green] curves), and $600\pi$ radians
(bottom two [red] curves), with the inverse square dependence
removed. (Unlike Fig.~\ref{cuspg}, a coordinate 
system, where $\phi = 0$ corresponds to the X-axis,
has been used -- hence the flat sections of the curves occur
at $1.5\pi$). The three straight lines plot the product, 
$\varepsilon \phi$ for the three distances (where $\varepsilon$ 
has been arbitrarily set to 0.05 
at $200\pi$, and drops as 1/distance as does that defined in 
Fig.~\ref{cuspg}.  
The topmost and bottommost curves represent the sum 
of the original curves and the lines (the same holds for the middle
curve, only with a smaller offset to the sum curve -- this curve
is also plotted in Figure \ref{xtry}).
}
\label{roundopsi}
\end{figure}

\begin{figure}[ht!] 
\centering
\includegraphics[height=15cm,width=19.344cm]{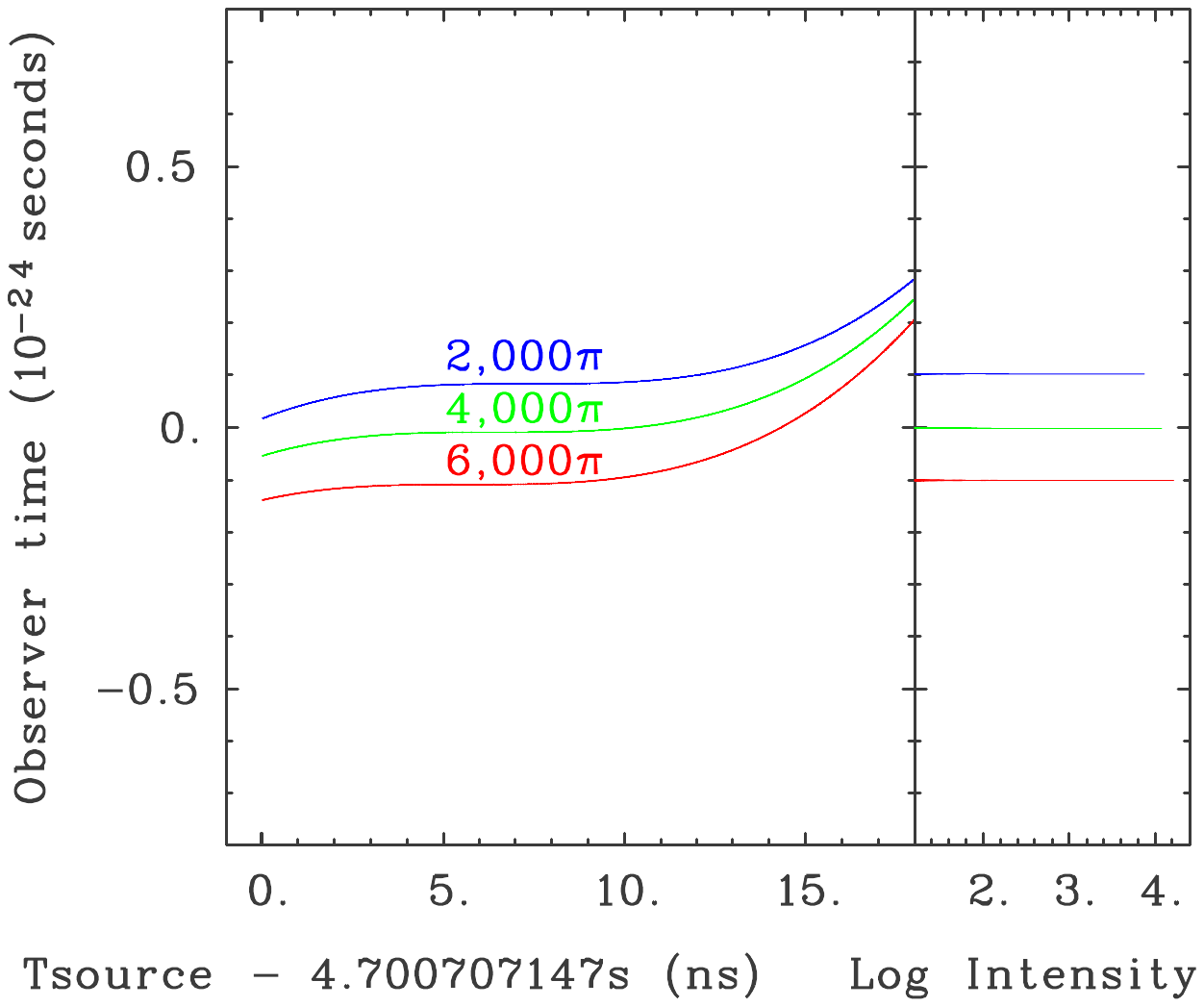}
\caption{Observer time as a function of Source time curves for an 
\textit{actual} pulsar with a period of 2$\pi$ and plasma at $R=5$, 
for three different distances, $2,000 \pi$ (top [blue] curve), 
$4,000 \pi$, (middle [green] curve), and $6,000 \pi$ radians
(bottom [red] curve), and, as with Figure \ref{roundopsi}, the 
inverse square dependence removed. (Unlike Fig.~\ref{cuspg}, a 
coordinate system, where $\phi = 0$ corresponds to the X-axis,
has been used -- hence the flat sections of the curves occur
at $1.5\pi$).  
}
\label{roundoga}
\end{figure}

We can see how an error in the linear term in $\phi$, call it 
$\varepsilon\phi$, where $\varepsilon$ is the small angular value from 
Fig.~\ref{cuspg}, affects the pulse profiles when included in the 
contributions, as has been done in Fig.~\ref{roundopsi} for a distance 
of $200 \pi$, with $\varepsilon = 0.05$ at this distance, and dropping
as 1/distance for the remaining $400 \pi$ and $600 \pi$.  The peak 
heights of the pulse profiles of this figure do, in fact, represent a 
1/distance law. The actual value for the linear error coefficient 
multiplying $\phi$ is orders of magnitude smaller, and numerically 
challenging to compute at sufficiently high resolution to show the 
effect, as we will see in Fig.~\ref{roundoga}. No matter how small 
this effect, however, it will produce a 1/distance law.

With no distance law incorporated into the three sum curves in the 
right hand frame of Fig.~\ref{roundopsi}, the more distant 
(red/lowermost) peak is 0.5 logarithmic units higher than the 
(blue/uppermost) peak which is three times closer, exactly compensating 
an inverse square law (when applied) to a 1/distance law. As long as 
the sampling over Observer time is kept sufficiently fine, and that 
over Source time is done likewise to maintain a sufficiently large 
sample to produce the pulse profiles in the right-hand frame, there is 
no contribution too small to produce this effect, again because the 
pulse profiles are otherwise infinitely narrow. Sixteen million plus 
points were needed, across the vertical of the right hand frame in F
Fig.~\ref{roundopsi}, to reveal the 1/distance law for a modulation 
of $0.001~\varepsilon(20\pi,40\pi,60\pi)~\phi$ (or $\varepsilon \phi$ 
for distances of $20,000 \pi$, $40,000 \pi$, and $60,000 \pi$). Half 
as many points led to an effect which matched the 1/distance law
poorly, whereas twice as many points led to a quantitatively 
excellent fit.

Calculations were also done for a pulsar of period $2\pi$ at 
distances of $200\pi$, $400\pi$, and $600\pi$ down to Source time 
range of $\sim$10 ns and increments $<10^{-15}$ s and an Observer 
time range of a small fraction of $10^{-25}$ s. By fine adjusting
the distances by parts per 10 billion, the centering effects of the 
peaks could be dithered. Although the responses at distances of 
$600\pi$ and $400\pi$ were essentially a ``dead heat'' after
compensating for the inverse square law, the response at $200\pi$ 
was down by 22\% -- the first noticeable manifestation of an 
inverse square law violation. By contrast, the result of 
calculations done with distances of $20\pi$, $40\pi$, and $60\pi$, 
produced the strongest signal, removing the inverse square 
dependence, at the smallest distance, $20\pi$. With a factor of 
fifty decrease in Source time range for the $100$'s of $\pi$'s 
calculations, and a similar factor of 10,000 in Observer time, 
the $\pm10^{-31}$ s resolution of this time in the 128-bit 
double precision becomes apparent. 

When doing these calculations,
it was important to use an Observer time resolution sufficiently
small to prevent contributions of one distance splitting into two 
neighboring pulse profile bins, which caused the results to 
vary much more widely. When adjusted, the small Observer time
range of the actual peaks would almost always fall into a single 
bin, with the pattern, of a fixed number of adjacent empty bins 
between those with (continuous) content, persisting for hundreds
of consecutive bins.

Further calculations were done for distances of $2,000 \pi$, 
$4,000 \pi$, and $6,000 \pi$. The Source time resolution was
$2.2 \times 10^{-14}$ s, and the Observer time resolution was
$1.9 \times 10^{-31}$ s, and the effects shown in 
Fig.~\ref{roundoga} are much more robust, producing about half
(on average, but more for the example plotted) of the 
logarithmic difference needed for a 1/distance law.
Calculations were also done for distances of $20,000\pi$, 
$40,000\pi$, and $60,000\pi$, with no real increase in
the range of responses, likely due to the larger distance
range causing the Observer time to hit the double 
precision limit for coarser resolutions.

The full effective timing advance/delay of the focused beam at 
distance, $f(\phi)$, as a function of $\phi$ at increasing 
distances becomes the sum of all terms,
\begin{equation}
\label{allvar}
f(\phi) = {\phi^3 \over 6} + \varepsilon \phi + \mathrm{Constant},
\end{equation}
and the constant is absorbed into the macroscopic distance. This
relation can be rewritten as a standard cubic equation,
\begin{equation}
\label{cubic}
\phi^3 + 6\varepsilon \phi - 6f = 0,
\end{equation}
which has the standard solution:
\begin{equation}
\label{solution}
\phi = (3f +(8\varepsilon^3 + 9f^2)^{1/2})^{1/3} + 
       (3f - (8\varepsilon^3 + 9f^2)^{1/2})^{1/3}.
\end{equation}
To get the maximum peak height we differentiate this wrt $f$ 
and set $f = 0$. The result is:
\begin{equation}
\label{derivative}
({\partial \phi} / {\partial f})_{f = 0} = 1/\varepsilon .
\end{equation}

Thus if, with an increase in distance, the slope of the linear term 
in $\phi$, $\varepsilon$, decreases by the same relative measure, 
then, in consequence, a greater range of the $R$-circle contributes 
to the same, narrow width of the pulse.  Starting with a very small 
range of contribution in $\phi$, this can continue until almost two 
radians of the R-circle contribute, when the changing polarization 
limits further contributions (see Section \ref{Pol}). Even when this 
happens, the same range of $\phi$ contributes to an increasingly 
narrow pulse width in Observer time as distance increases. So 
although the pulse height is restricted, beyond this point, to an 
inverse square law, the narrowing of the pulse represents that much 
more energy from its increasing range of Fourier components, thus 
the energy of the pulse still obeys a 1/distance law.  This is 
known as ``focusing in time.''  

More quantitatively, if we let $\delta$ represent a small interval/distance
within the pulse profile that the observer sees, then there will be 
some $\phi_{limit}$ which represents the contribution of source from
the $R$-circle whose energy falls within that $\delta$ for the observer:
\begin{equation}
\label{filimit}
\delta = {{\phi_{limit}}^3 \over 6} + \varepsilon \phi_{limit}, 
~~~~~\textrm{(in radians)}.
\end{equation}

Multiplying this equation by $R_{LC}$ restores the original units of distance, 
as in the first term of Eq. \ref{ro(RXY)}, the square root of distance squared.  
Then dividing by c converts this into an equation involving time.  It can then 
be converted into radians by multiplying by the rotation frequency, $\omega$.  
These three factors cancel exactly \textit{by definition}, so Eq.~\ref{filimit} 
involves radians, and needs no conversion.

Given that $\phi^3 / 6$ is small, we get $\phi_{limit}=\delta / \varepsilon$.
The resulting pulse profile will be a single sharp 
peak of a certain, narrow width, $\delta$, at least until $\phi_{limit}$ 
approaches $(6\delta)^{1/3}$.  Setting the two contributions equal, we get 
an equation for $\delta$ in terms of $\varepsilon$: 
$\delta=\sqrt{6}{\varepsilon}^{3/2}$, which gives the upper limit to 
$\delta$ above which the further 
contributions of source on the $R$-circle will be limited, unless $\delta$, 
and hence $\varepsilon$, are reduced, so that the 1/distance response can 
continue for greater ranges. We will continue to explore this in Section
\ref{AGN}.

\section{Further Considerations}
\label{Further}

The equation for $\rho$, now that we know $Y = 1/R$, and $Z(X,R)$, is:
\begin{equation}
\rho(X,R,\phi) = [(R\sin(\phi) - X)^2 + (R\cos(\phi) - 1/R)^2 
                  +(R^2 - 1)(X^2 - 1 + 1/R^2)]^{1/2}  .
\end{equation}
If we differentiate this wrt time, and set $\phi$ to 0, it simplifies
drastically to:
\begin{equation}
\rho(X, R, \phi = 0) = XR.
\end{equation}
And if we use the fact that $d\rho/dt = \textrm{c}$, it follows that:
\begin{equation}
{dX \over dt}    = \textrm{c}/R,
\end{equation}
and this is even true (on average) between $X = 0$ and $\sqrt{1 -1/R^2}$.

The velocity of the focus, on the other hand, is infinite and diverging in 
the Z direction at $Z = 0$, and becomes finite at $Z \neq 0$, but remains
greater than c, even if ever so slightly, for the rest of its path to 
infinity. Causality is not violated since the path is not a straight line.
Also, because the path of the focus is not radial from the contributing 
currents on the R-circle, the elements making up the focus represent 
parts of their spherical emission which change continuously as the
focus moves outward.
But although these paths also \textit{curve} over toward $+X$, 
once $Z$ is no longer 0, the progressive narrowing of the vCc's caused by 
the extreme deceleration prevents any secondary focus from forming.  The 
focus we've discussed is the \textit{only} focus for circularly 
supraluminal excitation.

We can derive the distance scale of change for the angle, $\varepsilon$ (not
restricted to the X-Y plane) by taking the cross product of the vector to the
source of excitations, $(0,R,0)$, with the $\rho$ vector: 
$(X,1/R-R,\sqrt{R^2 - 1}\sqrt{X^2 - 1 + 1/R^2})$, and then dividing by the 
moduli of the two vectors, namely $R$ and $RX$ respectively. This quotient 
will yield the value of $\cos(\varepsilon)$, which for great distances will 
be slightly less than unity. 
The cross product is $(R\sqrt{R^2 -1}\sqrt{X^2 -1 + 1/R^2}, 0 ,-RX)$, and the
quotient is:
\begin{equation}
\cos(\varepsilon) = 1 - {1 \over X^2}(1 - 1/R^2)^2 .
\end{equation}
At great distances we can estimate $\varepsilon$:
\begin{equation}
\varepsilon = {{R - 1/R} \over {\rho}}; ~\textrm{where}~\rho = RX .
\end{equation}
as we might have guessed.

So far we have discussed the situation involving just a moving point source of
emission, whereas in the real world \textit{volume} sources are involved, and
issues such as plasma screening rear their ugly heads. However, pulsars really
do obey the 1/distance law we've discussed here, so there must be some way that
pulsar emission can act like a lot of point sources, possibly by field lines 
bunching together, as suggested by [\onlinecite{Contopoulos}]. Temporal 
structure as short as 0.4 ns [\onlinecite{HE07}] has been observed in the 
giant pulses of the Crab pulsar, yielding a range in timescale close to 80 
million (33 ms/0.4 ns). However, we will see in Section \ref{AGN} evidence 
for a range close to the \textit{square} of this factor.

\section{Polarization}
\label{Pol}

The polarization of all sharp pulsar peaks observed in the Universe 
swings from nearly $+\pi/2$ to $-\pi/2$ across the pulse (or vice-versa 
-- see, e.g., Fig.~2 of [\onlinecite{MP87}]), and this can be understood 
from Figs.~\ref{cuspg} and \ref{xtry}, as the source time of the pulse 
moves across a macroscopic range.

For any given electric field element in the X - Y plane, $V$, with 
components $(V_x,V_y,0)$, there is a polarization field, $W = 
(W_x,W_y,W_z)$, for the direction, $\rho = (\rho_x,\rho_y,\rho_z)$, which 
has to be perpendicular to $\rho$, \textit{and} lie in the $\rho-V$ plane.  
So we have ${W \bullet \rho} = 0$, and $W \bullet (V \otimes \rho) = 0$.  
There is a penalty for the angle between $V$ and $W$, which is just the 
dot product between the two, once $W$ has been determined, as described 
below.

\begin{figure}[ht!] 
\centering
\includegraphics[height=16cm,width=20.6336cm,
trim={1cm 2cm 1cm 1.1cm}]{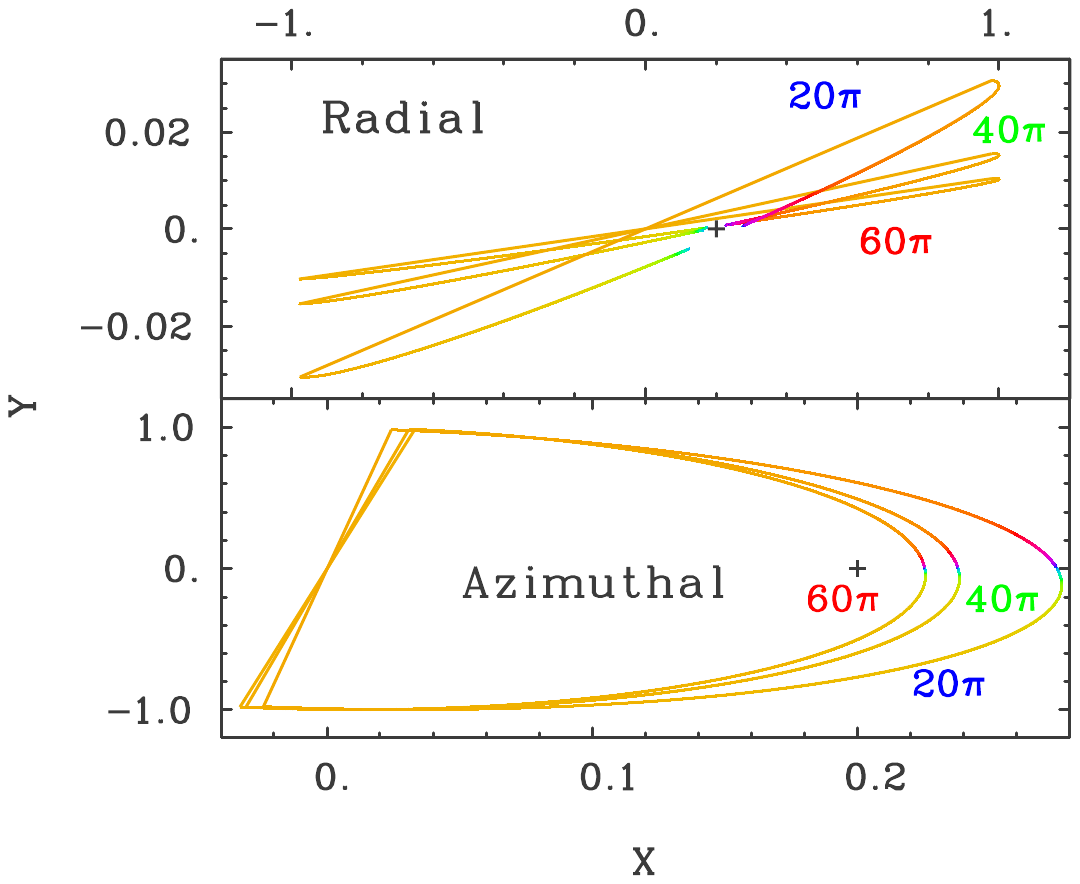}
\caption{
The X and Y offsets from the Z axis for a unit vector derived from 
the cross product of the polarization vector with $\rho$ at phase
$3\pi/2$ (in a right-handed coordinate system, i.e., ``approaching''),
for a pulsar with a period of $2 \pi$ seconds and plasma at $R = 5$, 
for distances of $20 \pi$, $40 \pi$, and $60 \pi$ light-radians.  
Loci for polarizations resulting from an electric field vector of 
$(\cos{\phi}, \sin{\phi})$ (radial), and $(-\sin{\phi},\cos{\phi})$ 
(azimuthal) are plotted in the bottom and top frames.  
The color coding of the loci is the same as for Fig.~\ref{xtry}.  
}
\label{z5m}
\end{figure}

For a radial electric field, $V = (\cos({\phi}), \sin({\phi}))$ (in 
a right-handed coordinate system with $\phi = 0$ for the +X axis), 
and using $W \bullet \rho = 0$ to substitute for $W_x$, we get:
\begin{equation}
\label{radial}
W_z =  W_y \rho_z(V_x + V_y \rho_y/\rho_x)/(V_x \rho_y - 
V_y({{\rho_z}^2}/\rho_x + \rho_x)).
\end{equation}
By using $({W_x}^2 + {W_y}^2 + {W_z}^2) = 1$, all three components
of $W$ are determined.  
For an azimuthal electric field, $V = (-\sin(\phi),\cos({\phi}))$,
when substituting for $W_y$, we get
the same equation with subscripts `x' and `y' exchanged. The loci of 
these $W$s,
in the plane of the $Y$ direction and that perpendicular to the 
X-Z asymptote ($Z = \sqrt{R^2 - 1} X$), are simple to describe.  



The results for both the radial and azimuthal cases are shown in 
Fig.~\ref{z5m}. Polarization offsets with magnitudes inversely 
proportional to distance are present in both orientations. 
More importantly, however, is the fact that the angle of polarization
swings \textit{drastically} during the period of maximum response, as is
both predicted and observed. 
Although the residual offset of the points from (0.2, 0.) in the lower
frame, and the residual slope of the curves in the upper frame approaches
0., a residual slope does persist in the linear parts of the curves in 
the lower frame.

In addition, when constructing a closer ns-scale Source time counterpart 
to Fig.~\ref{roundopsi}, points at distances of $10 \pi$, $20\pi$, and 
$30 \pi$ would have to be retarded in phase by 0.02122, 0.03183, and
0.06364 radians, respectively, in order to have their inflections 
coincide, on the ns scale, with $1.5 \pi$, an effect presumably due to 
approximation errors and/or higher order than cubic terms of the 
expansion of Observer time in powers of $\phi$, which also have a 
1/distance dependence. 

\begin{figure}[ht!] 
\centering
\includegraphics[height=12cm,width=15.6cm,trim={1cm 2cm 3cm 1.1cm}]
{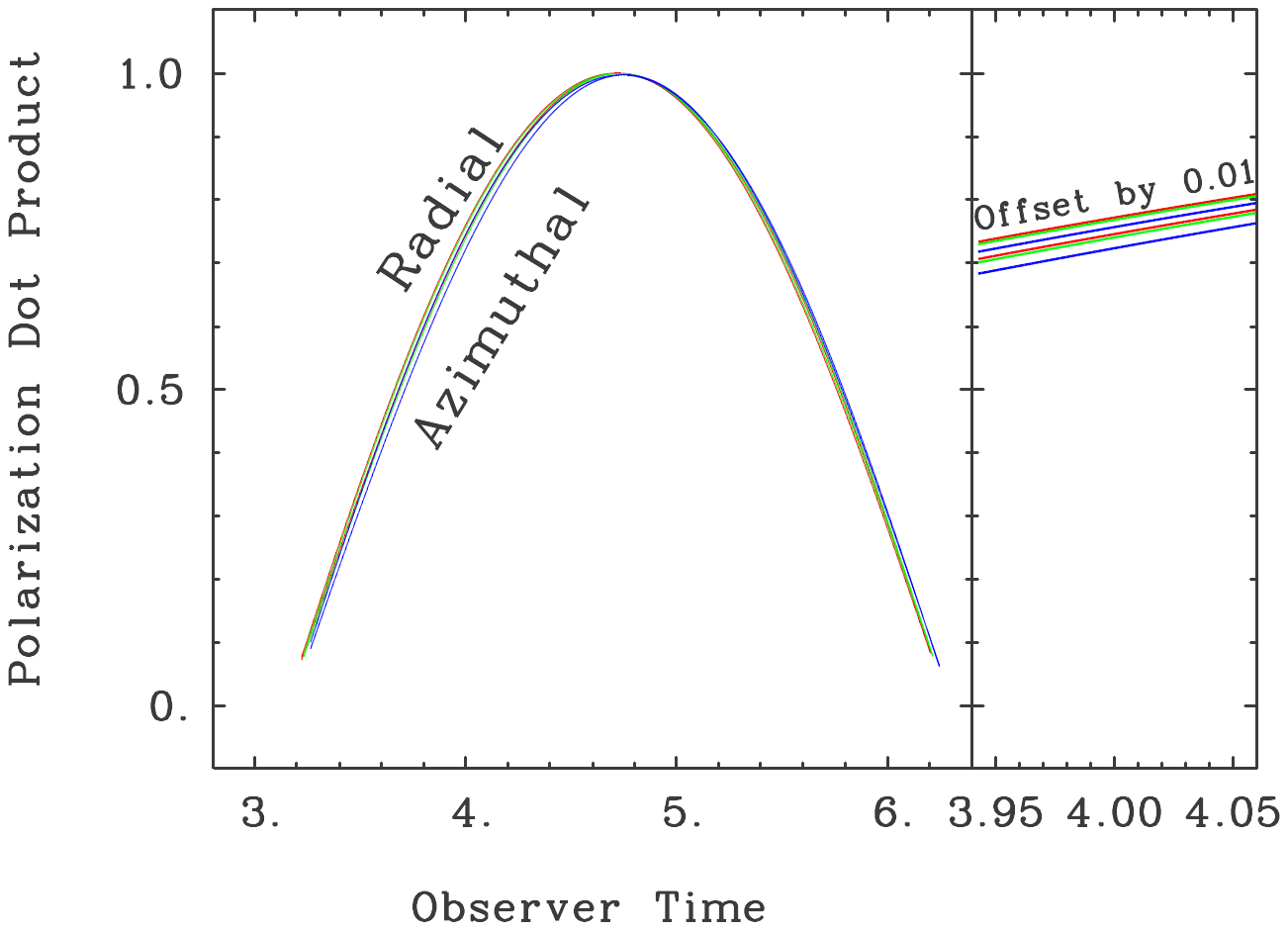}
\caption{The dot products of polarizations, resulting from radial and 
azimuthal electrical field vectors on an annulus at $R = 5$, with their 
counterparts at $\phi = 0$ (using the convention of Fig.~\ref{cuspg}), 
for distances of $20 \pi$, $40 \pi$ and $60 \pi$. (Right frame) A
close up of a small section of the curve in the left frame, with the
three radial dot product paths offset up by 0.01.
}
\label{dot}
\end{figure}

Finally, the dot products of the polarization vectors with their $\phi=0$
(as in Fig.~\ref{cuspg}) counterparts are shown in Fig.~\ref{dot}, which 
confirms they rotate as time progresses, as well as small differences 
between distances and electric field orientations. In addition, 
Fig.~\ref{dot} confirms the above statement that the points of inflection 
in Observer time as a function of Source time occur at slightly different 
phases at different distances. 

These two figures show, at least for the $R = 5$ case, that the 
polarization vector has significant components of rotation, other than 
the overall rotation with phase, at $\phi = 0$ (as in Fig.~\ref{cuspg}), 
for azimuthal and radial surface electric fields, with a rate which 
is inversely proportional to $\rho$.  


\section{Implications for known pulsars and supernovae}
\label{MI}
\subsection{Pulsars}
\label{pulsars}

At least half of all known pulsars --- those whose pulse profiles 
consist of a single, sharp pulse of width less than $\sim$3\% of 
their periods --- have been discovered because their rotation axes 
are oriented nearly perpendicular to the line of sight to the Earth.  
Isolated neutron stars must gravitationally concentrate interstellar 
plasma in order to emit radiation (via cyclotron emission and/or 
strong plasma turbulence) mostly in a location \textit{outside} of 
the light cylinder, because this region marks the beginning of 
supraluminally generated focused radiation.  

And the place 
\textit{in this region} where plasma is most highly concentrated 
is \textit{just} outside the light cylinder. For isolated pulsars, 
this emission occurs at\cite{M12,Weisskopf} $R=1/\cos{4^{\circ}} = 
1.00244$.  This means that 
the 
speed of the excitation, $v$, is just slightly greater than the speed 
of light, c, so that $\arcsin(\mathrm{c}/v)$ is almost $90^{\circ}$ 
--- the cones of favored emission are very open, and the favored 
emission is very nearly equatorial. This may have observational
consequences.

For example, no sustained central source has been detected in any 
nearby modern SN except 1986J, and, by now, even that source does 
not appear to be consistent with a strongly-magnetized pulsar 
remnant.\cite{BB17} It is also the \textit{only} 
one viewed in the center of an edge-on galaxy, NGC 0891.  If the 
rotation axis of any pulsar remnant is perpendicular
to the NGC 0891 galactic plane, then we are close to its
favored direction of its emission (4$^{\circ}$), 
raising the possibility that the 
visibility of the central source in 1986J was not entirely 
fortuitous. The larger the star, the more likely its angular 
momentum aligns with that of its host galaxy, and all the more 
so for binary mergers.

The implications of a 1/distance law for pulsar emission, even 
restricted to \textit{exactly} two opposite spin latitudes, are 
far-reaching (to coin a phrase).  It is responsible for the 
continued success of a long string of pulsar searches made over the 
last several decades. It may be responsible for the utraluminous 
X-ray source found in NGC 5907.\cite{ULX} Although currently, 
no radio pulsar has been discovered in any galaxy beyond the Large 
and Small Magellanic clouds, the Square Kilometer Array may change 
that.\cite{ska} The nearby galaxy in Andromeda, M31, and another 
nearby spiral, M33, may yield many pulsars each. 

\begin{figure}[ht!]
\centering
\includegraphics[height=14.2078cm,width=20.6336cm,trim={1cm 2cm 0cm 
1.1cm}]{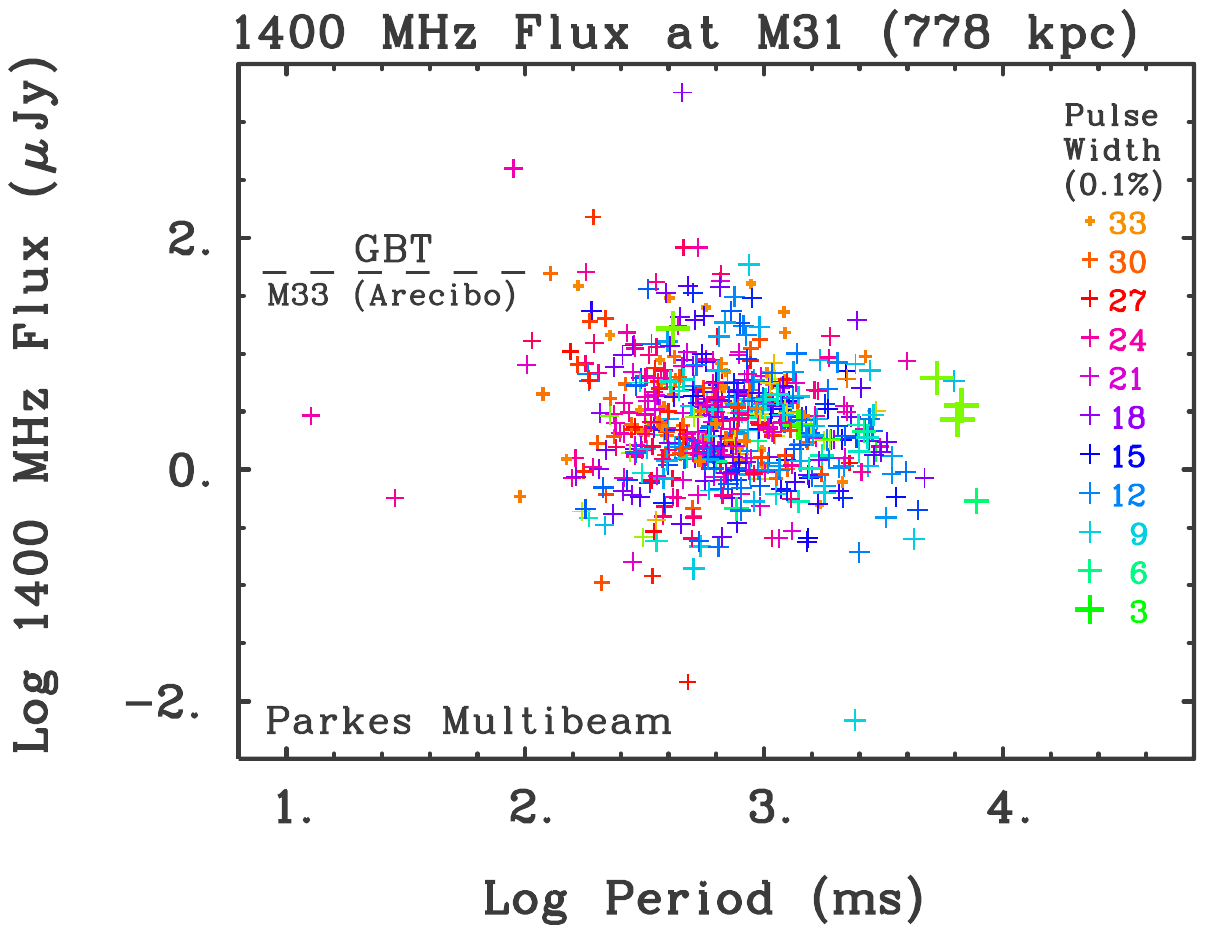}
\caption{
The 1400 MHz fluxes of the 577 Parkes Multibeam pulsars, with pulse 
FWHMa$<$0.034 cycles, when placed at the distance to M31 (778 kpc), 
assuming that all obey a strict 1/distance law (and ignoring their 
individual placings in the Milky Way). The dashed line at the left 
represents the approximate limit of the NRAO Green Bank Telescope 
for M31, and that of the (now defunct) Arecibo 300-meter dish for 
another spiral galaxy, M33, at $\sim$900 kpc (at $41^{\circ}$ 
declination, M31 is out of reach from Arecibo, but like M33 makes a 
good target for FAST\cite{FAST}).
}
\label{s1400}
\end{figure}

However, we are not there yet, and care must be taken when making 
estimates of how far we can go presently.  For example, 
[\onlinecite{Rubio-Herrera}] determined that, if the ``most 
luminous'' pulsar in the Galaxy, PSR B1302-64, were placed in M31, 
it would need to be five times as luminous in order to be 
detected in a survey using the 12, 25-meter antennae of the 
Westerbork Synthesis Radio Array. Since M31, at 778 kpc, is 64.35 
times farther away than B1302 is from the Earth ($\sim$12 kpc), it 
can be inferred that B1302 was robustly detected at 13 times the 
threshold level of the survey.

However, B1302 is only the most luminous pulsar assuming that its 
flux obeys {\it an inverse square law within the Milky Way!} With a 
571 ms period and a pulse width\cite{scat} of 19 ms, B1302 is almost 
certainly near a 1/distance law, and thus its luminosity is 
overestimated.   

Figure \ref{s1400} plots what the 1400 MHz fluxes would be at the 
778 kpc to M31 for the 577 Parkes Multibeam pulsars whose pulse FWHM 
is less than 0.034 cycles (so that B1302-64 is included). On this 
graph the flux of B1302 at M31 is only 25 $\mu$Jy, lower than 21 
other pulsars.\cite{W1400} With 310/1.8 mJy at 4.5/778 kpc, the 
sharply pulsed (0.018) B1641-45 is the clear winner at M31. The bulk 
of the population lies between 1 and 10 $\mu$Jy.

For the Magellanic Clouds themselves, the 59.7 kpc-distant Small 
cloud has four known pulsars, with pulse widths of 0.0300, 0.0216, 
0.0189, and 0.0138 cycles, while the 49.7 kpc-distant Large cloud 
has 12, with two young, strongly magnetized pulsars at 62 (X-rays 
only so far), and 20 Hz, both of which were discovered in the X-ray 
band. It also has 10 more exclusively radio pulsars, with pulse 
widths of 0.0818, 0.0326, and 0.0246 cycles, and the remaining seven 
with still narrower peaks --- like the SMC a narrower sample than 
the the full PKSMB --- lending support to the pulsar model under 
discussion.\cite{H98,Narrow}

The 20 Hz pulsar in the LMC (B0540-69.3) has a double peak spanning 
one third of a cycle (see the middle pulse profile in 
Fig.~\ref{rounda}), characteristic of a viewpoint which is more 
equatorial than optimum ($4^{\circ}$). The shock(s) of the Chandra 
image of this pulsar appear as a straight   
line\cite{Serafimovich,ch04} consistent with our view being 
extremely close to equatorial, again lending support to the 
supraluminal excitation model of pulsar emission.
  
The 62 Hz pulsar in the LMC, J0537-6910,\cite{M06} has a marginally 
narrow pulse ($\sim$0.1 cycles), which appears to be a close double. 
Thus our view is farther from the equatorial than it is for B0540, 
and closer to optimum. Its nearly-aligned magnetic field is drifting 
toward its spin equator by about a meter per year, but the field 
\textit{strength}, rather than effective dipole, is what matters for 
cyclotron radiation. In the case of J0537, its field is still strong 
enough, at its close-in light cylinder, to produce X-rays, but few 
optical photons because the cyclotron process can not produce 
subharmonics, still another test that the model passes, and few 
others do.

Similar strongly magnetized pulsars, such as that in the Crab Nebula 
(30 Hz), may drive winds which move material through positions of 
favored supraluminal excitations relative to an observer. The 
Chandra image of the shocks near the Crab 
pulsar\cite{ch04,Weisskopf} show an ellipticity consistent with our 
view being 29$^{\circ}$ off the rotational plane, an orientation 
which would produce the weakest (top) pulse profile shown in 
Fig.~\ref{rounda} --- i.e., the interpulse. The very sharp main 
pulse (a 1.1 ms FWHM) could be the result of emission from the wind 
where its material passes through the favored location at $R = 
1/\cos(29^{\circ})=1.143.$ The absence of the GHz emission bands, 
seen in the interpulse, is consistent with the less homogeneous 
environment from which this feature must arise.

Finally, the pulse profiles recorded for the periodic 2.14 ms 
signal\cite{M00,M00c} from SN1987A were rarely anywhere near as 
sharp as are at least half of all radio pulsars, with the smallest 
pulse widths at no less than 10\%. Thus it was difficult to 
reconcile this pulsation with a real source at 49.7 kpc, with a 
violation of the inverse square law apparently ruled out. The 
answer, in this case, is that an otherwise sharp pulse is 
broadened by the phase variation of the \textit{precession}, for 
which there is evidence in all of the results sufficiently 
significant to reveal it.

\begin{figure}[ht!]
\centering
\includegraphics[height=12.3485cm,width=19.344cm, trim={1cm 2cm 
-2cm 1.1cm}]{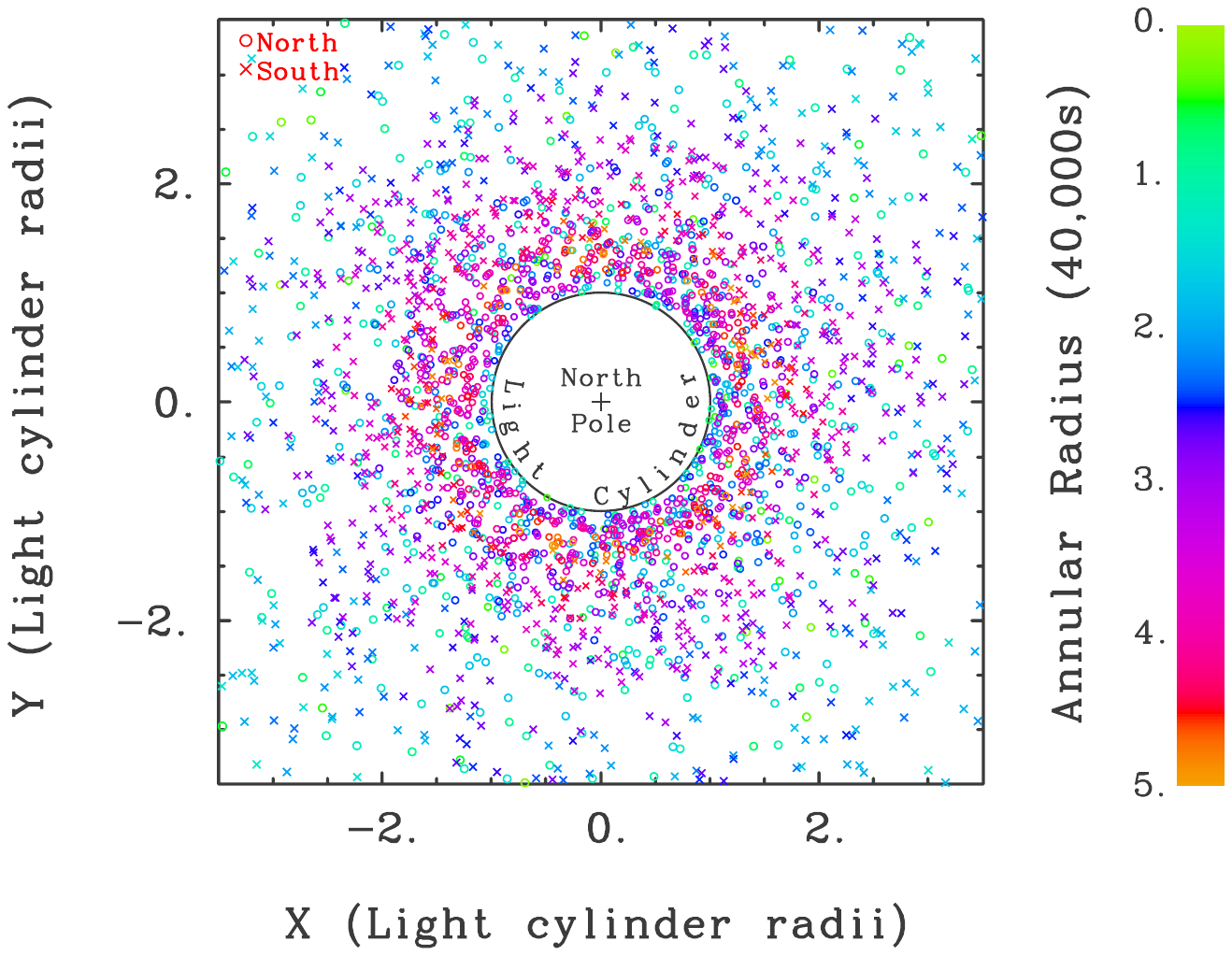}
\caption{
The emergence locations (X-Y, near the northern rotational pole) 
for the 2,655 focused beams which fall within the (X=$\pm$3.5,
Y=$\pm$3.5) radians of the frame out of 3,930 annuli of 
polarization currents within a post-core-collapse star of 400,000 
radians diameter (a 19.08 million km radius for a 500 Hz pulsar, 
and a 190.8 million km radius for a 50 Hz pulsar). The annuli 
are spaced in polar depth and radius by about 1,000 radians.  
Phase lag has not been accounted for, making the azimuthal locations 
for the emergences essentially random. The `o' symbols mark the 
emergences from annuli in the stellar northern hemisphere, and 
`x' symbols for those from annuli in its southern hemisphere.
}
\label{SN}
\end{figure}

\subsection{Supernova disruption}
\label{SNdis}

The 1/distance law is essential to pulsar-driven supernova 
disruption --- the progenitor star radii are in the range of 1 
-- 40,000 light-periods ($2\pi R_{\mathrm{LC}}$/c), whether 
white dwarf -- white dwarf or blue supergiant, core-merger,
$\sim$~500 Hz pulsars, or solitary red supergiant, $\sim$~50 
Hz pulsars (see Fig.~\ref{SN}). Annuli of polarization 
currents close to the poles, as well as annuli at all but 
the smallest radii deeper from the poles, all contribute to 
beams which are concentrated at the poles.

\begin{figure}[ht!]
\centering
\includegraphics[height=14.6114cm,width=19.344cm, trim={1cm 
2cm -0.6cm 1.1cm}]{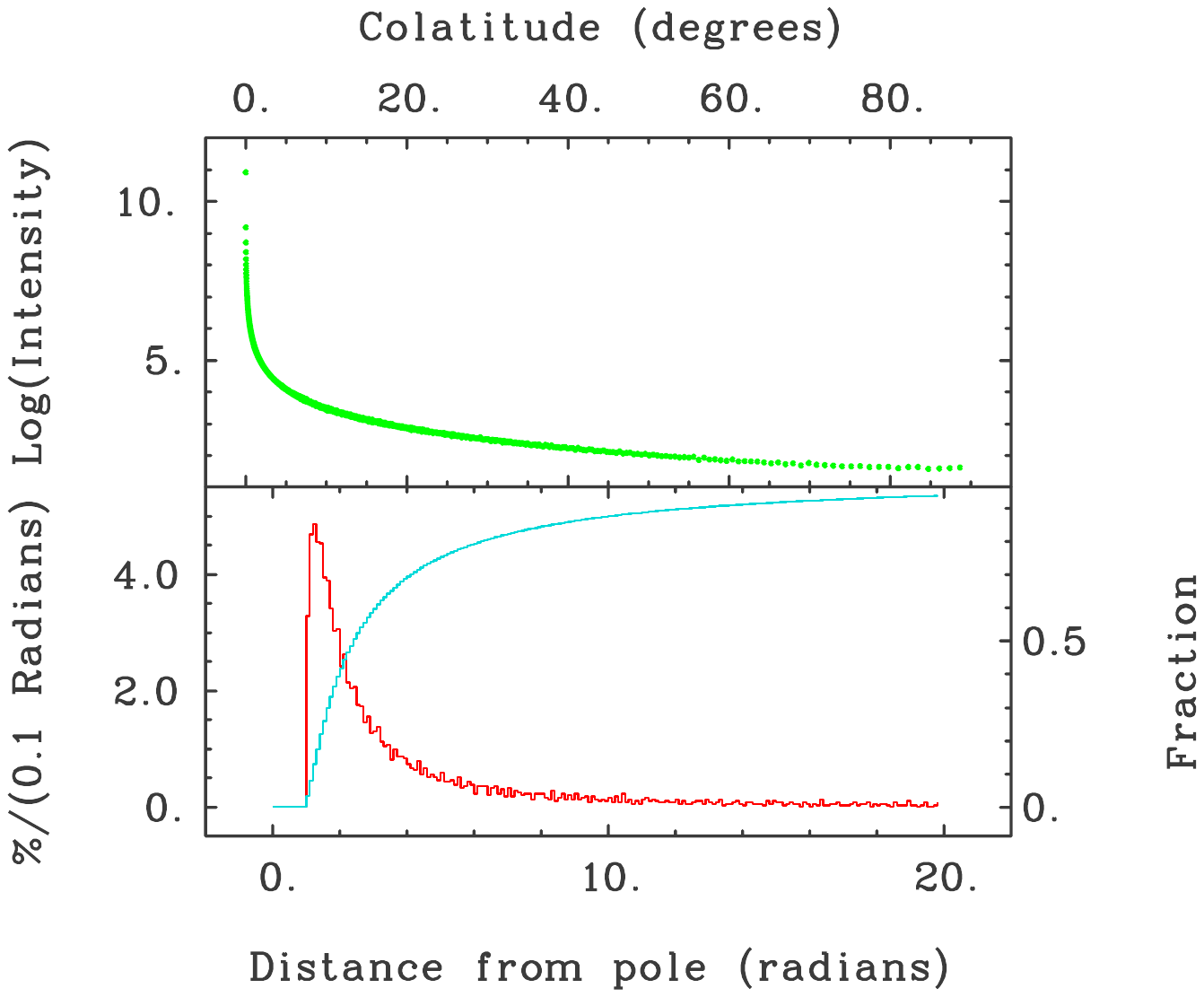}
\caption{
(Lower frame, left hand vertical scale) The percentage of 3,930 
cusp emergences (red) histogrammed in steps of 0.1 radians 
projected polar radius. (Lower frame, right hand vertical 
scale) The cumulative fraction of emergences (cyan) vs projected 
polar radius. (Upper frame,left hand vertical scale)  The log 
of the number of cusp emergences, at all azimuths, per degree 
of colatitude (upper horizontal scale). The highest resolution 
of 0.01 degrees corresponds to 35 radians in the horizontal 
scale of the lower frame.
}
\label{coll}
\end{figure}

Thus the jets driven from this radiation are polar, 
emerge within \textit{minutes} of core collapse, and can be 
collimated to a high degree,\cite{M12} easily one part in 
10,000 (see Figs.~\ref{SN} and \ref{coll}), and still more for 
larger stars, making this one of the most anisotropic known 
processes in the Universe. 

The advantage in $1/r^3$ dipole strength for the near-axially 
induced/smaller radius, $r$/more equatorially erupting beams, 
over the less axial/larger $r$/more polar beams is not as great 
as might be expected. The contributions to the latter grow 
larger with $r$, are boosted by a centripetal acceleration 
which also grows with $r$, and are focused onto the smaller 
polar rather than the larger equatorial regions.  In the end,
the excitations/beams to both regions must propagate out of 
the stellar core and at least the rest of the stellar radius, 
with the paths from the annuli of greater radii avoiding more 
and more of the stellar core as they penetrate to the poles.  

Thus a plot of the number of beam emergences vs colatitude 
(upper frame of Fig.~\ref{coll}) may be a decent measure of 
anisotropy, although no adjustment has been made for the 
differing stellar densities at the locations of the polarization 
currents. However, pulsar precession [\onlinecite{M00,M00c}] 
limits the effective anisotropy to about a factor of 100, arguably
even less than the collimation ($R \le 210$) implied by ``Bright 
Source 2'' of [\onlinecite{NP99}], which was roughly the same 
physical size, but much more distant from SN 1987A. For stars 
with circumstellar material, a meandering pencil beam cuts through 
this gas, producing copious amounts of radiation along the way.  
The amount of radiation produced may be comparable to that which 
emerges from the poles.

This is not unexpected, considering the SN 1987A ``Mystery 
Spot'' or ``Bright Spot 1'' -- a  feature observed on days 30 
and 38 by [\onlinecite{Nis}], and day 50 by [\onlinecite{MMM}], 
to be consistent in magnitude ($\sim$6\% of the optical flux 
of the SN proper), position angle ($194^{\circ}$), and, with a 
reasonable extrapolation, offset (45, 60, and 74 milli-arc
seconds, or a few light-weeks south by southeast of the SN 
proper) -- and its anisotropy,\cite{WWH} visible in the remnant 
now for the last few decades. For large stars either the 
\textit{core merger process} within the common envelope,\cite{M18} 
or the reduction of the core's moment of inertia through 
progressive nucleosynthesis, initiates\cite{M} this anisotropy 
even \textit{before} core collapse.  

Early spectra of SN 1987A indicate ``bright sources underlying 
diffuse material,'' not unlike glowing pyrotechnic embers viewed 
through intervening smoke. Much, but certainly not all, of 
this underlying luminosity must come from the visible South 
pole.  The rest must originate from the effect of the polar 
jets penetrating their overlaying circumstellar material,
which, given the 9-day delay (Fig.~3 of [\onlinecite{M12}]) 
and the 75$^{\circ}$ orientation, indicates such 
material at locations light-weeks above the poles.

Following a small, 1-day increase at day 7.8 due to the UV 
flash hitting, but not penetrating deeply into, the 
circumstellar material, the luminosity from the jet penetration 
continues to ramp up beyond day 9. Following an additional 
small spike at day 20 (most prominently in U, R, and I),
when what remains of the UV flash breaks out into a relative 
clearing, a decrement occurs over days 20 to 21, when the 
particle jet enters the same clearing. The luminosity then 
continues to ramp up after after this one-day delay with 
almost the \textit{same} slope (more supra-polar material).  
Figure 3 of [\onlinecite{M12}] shows that by day 25 this 
luminosity ramp was only a magnitude fainter than the 
minimum of $m_V = 4.5$ at day 6.8. Since the ramp 
back-extrapolates\cite{M12} to the same value at day 7.8,
as the minimum at day 6.8, it appears that the stellar 
luminosity is not rapidly increasing at that time (due, 
for example, to production of $^{56}$Ni), nor would it 
also be expected to increase due to the disruption 
mechanism acting on the South pole, since precession will 
dominate any tiny changes expected in the beam/jet 
collimation with time.  
Thus as much as 28\% of the total SN 1987A luminosity at 
day 25 was due to polar jets(s)/beams(s) interacting with 
circumstellar material light weeks away from the star.

The luminosity continues to rise progressively more 
slowly to a peak and turns over after day 86,\cite{HS90} 
when much of the circumpolar material has been driven away, 
so that less of this was engaged by the jet(s). By this 
time the V flux had increased by a factor of 5.4.  So
although the ``Mystery spot'' at days 30, 38, and 50 amounts 
to only 8\% of the total light in H$\alpha$, it may be that 
diffuse, or out-of-band H$\alpha$ luminosity from the jet/polar 
circumstellar material is still greater. Not until day 
120 when the slope break occurs indicating $^{56}
$Co decay, can we be sure that the luminosity of the star 
proper totally dominates.

The early details did not show up in the B band because the flux 
was still falling from the UV flash until day 20, but after that,
follows the V curve closely. This means that much of the
luminosity near the peak of the SN 1987A light curve was due to
the jet(s) penetrating the circumstellar material, a flux whose 
intensity \textit{and its time of maximum} will both vary with the 
angle between the rotation axis and the line of sight to the Earth.

The circumstellar material expelled by SNe Ia progenitors will 
differ, slower than the 0.957 c of that in SN 1987A, and perhaps 
with less of an initial gap/mean depth than 1.5/2.5 light-weeks, 
but given that these are also merging binaries, as was SN 1987A, 
that material will still be out there, as polarization 
measurements confirm.\cite{Kasen03} Even though the rate of 
$^{56}$Ni production will be greater, in general the light 
curves peak earlier, so the interaction flux 
will still be an important fraction of the total, and may actually 
equal or exceed the relative fraction in SN 1987A because of the 
higher atomic numbers of C and O.  

Thus using these objects as 
standard candles is likely to be an exercise in futility, unless 
we can observe the early development, as proposed in 
[\onlinecite{M18}], in many such, nearby SNe, and somehow find 
observables still measurable in the distant sample that can 
unravel all the key parameters.

\section{Distance effects in pulsars, GRBs, and AGN jets}
\label{AGN}

For isolated pulsars, from earlier, emission occurs just outside of 
the light cylinder at $R=1/\cos{4^{\circ}} = 1.00244$, thus the scale 
of decrease of $\varepsilon$ is 0.0048, and its scale of decline for 
a 1 Hz pulsar, with a 47,700 km $R_{\mathrm{LC}}$, is 232 km. The 
spin rates of the 14 exclusively radio pulsars in the Magellanic 
Clouds range from 0.55 to 4.1 Hz. The implication of such pulsars in 
the LMC and SMC at 50 and 60 kpc, which the 1/distance law would 
place between 3.2 to 20.4 $\mu$Jy at the distance to M31 and not that 
far from the middle of the distribution shown in Fig.~\ref{s1400}, is 
that the scaling with $\varepsilon$ is still unrestricted by 
$\phi_{limit}$ over a factor of $6.6 \times 10^{15}$ in distance.  

\begin{figure}[ht!]
\centering
\includegraphics[height=14.cm,width=18.0544cm]{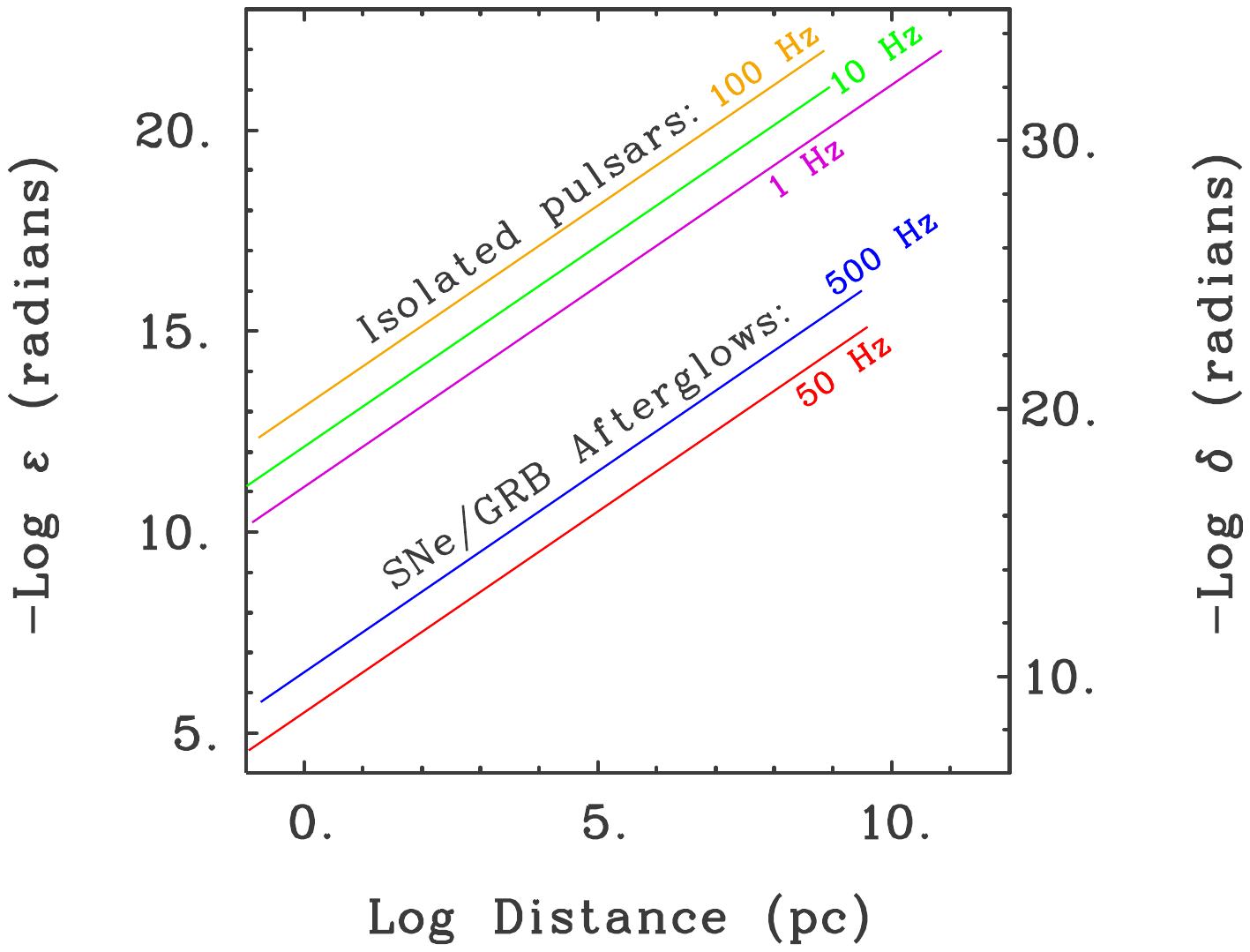}
\caption{
Left vertical scale: the -log of the angle, $\varepsilon$, from 
Fig.~\ref{oneovrppb4} as a function of the log distance (bottom 
horizontal scales), and the log of the angle, $\delta$ (right 
vertical scale), representing the upper limit for 1, 10, and 100 
Hz isolated pulsars, and disruption of red (50 Hz) or blue (500 Hz) 
supergiants, whose cores have collapsed to neutron stars.
}
\label{delta}
\end{figure}

For gamma-ray-burst afterglows, $R = 100,000$ times $R_{LC}$
amounts to $\sim$10$^7$ km for blue supergiants, and $\sim$10$^8$
km for red supergiants, but maybe only $R=100$ or $10^4$ km for 
mergers of two white dwarfs (a possible source of short gamma-ray 
bursts and their afterglows). To see GRB afterglows from these 
processes at 4 Gpc, the scaling with $\varepsilon$ must hold up 
for a factor of $1.3\times 10^{15}$ for red giants, ten times that 
for blue supergiants, and 10 million times
that for white dwarf mergers (see Fig.~\ref{delta}). 

It is not unreasonable to think that scaling with $\varepsilon$ will 
hold to another factor of 10 in distance, from $\sim$55 kpc to the 
778 kpc of M31 and the 898 kpc to M33, and that pulsars will be 
detected in these, even if the sampling in time is not fine enough 
to exploit any pulse narrowing (see three paragraphs below). Another 
conclusion which can be drawn from all this is that $\delta$ can be 
immeasurably small. Figure \ref{delta} shows the $\delta$ values for 
1, 10, and 100 Hz isolated pulsars, and the red and blue supergiants 
which produce GRBs and afterglows from their supernovae.

From Fig.~\ref{delta} and some work with a straightedge, we can see 
that the limiting pulse width ($\delta$) of a 1 Hz pulsar at 
$\sim$100 pc is $10^{-20}$ radians, or $\sim$1.6$\times 10^{-21}$ s. 
This rate is consistent with 1 ns after $\sim$4,900 km, and $\sim$20 
ns at one characteristic distance for the decrease of $\varepsilon$ 
(232 km).

When the coherent range does reach $\phi_{limit}$, and then the 
distance increases by some factor, $\alpha$, decreasing $\varepsilon$ 
by the same factor, the range of phases which fall below the limiting 
$\phi^3/6$ curve falls by a factor of $\sqrt{\alpha}$. From our 
discussion near the end of Section \ref{violate}, we learned that
the more limited source phase range has made 
the resulting pulse a factor of $\sim\sqrt{\alpha}$ narrower, which 
represents more \textit{energy} by the same factor because of the
increased bandwidth of the Fourier components of the narrower pulse,
and sensible detection algorithms can exploit this. So even here,
the effective distance law is still better than inverse square, while 
for less distant sources, every factor increase in distance causes
that much more source within the \textit{same} small time range,
thus the 1/distance law applies.

An interesting possibility is the observation of Crab-like optical 
pulsars at great distances. Calculating $R_{\mathrm{LC}} \times 
(R-1/R)$ for a 25 Hz pulsar which produces a sharp pulse by pushing 
material through a typical observer inclination-determined $R$ of 
$2/\sqrt{3} = 1.155$, or $60^{\circ}$ (30$^{\circ}$ off the 
rotational plane), gives a scaling length for $\varepsilon$ of 551 
km. It takes $5.6 \times 10^{10}$ of these to make just one parsec. 
Thus $\varepsilon$ is $30^{\circ}/(5.6 \times 10^{10}) = 
9.3 \times 10^{-12}$ radians, and $\delta = 7 \times 10^{-17}$ 
radians, or a whopping 1.3 $\mathrm{\AA}$ at $R_{\mathrm{LC}}$, or 
$4.3 \times 10^{-19}$ s. At this point, quantum effects may take over 
for all emission wavelengths longer than very soft X-rays, including 
the radio through the ultraviolet.

If a rotating, magnetized body can drag its magnetic field 
supraluminally across plasma, so can an orbiting, magnetized body, 
near the supermassive black hole in an AGN, and certainly any 
orbiting binary system. The focused radiation, which first moves in 
the Z direction (recall that $Z = \pm \sqrt{R^2 - 1} \sqrt{X^2 - 1 
+ 1/R^2}, ~\mathrm{for}~X = \sqrt{1 - 1/R^2}~+ $ `a small value,' 
starts completely vertically) is ideal for forming jets perpendicular 
to the orbital plane/thin accretion disk. Such jets will not be 
nearly as highly collimated as pulsar-driven jets, but the 
observational constraints only demand 1 part in 100 collimation (and 
the occasional magnetic field line advected into the jet from 
material in the accretion disk).

Since magnetic field lines can not thread into uncharged black holes, 
they must somehow adjust when two of them are co-orbiting prior to 
merging, in a manner that ordinary matter can not mimic. Thus it is 
possible that this situation will also result in magnetic fields 
moving supraluminally across matter, whose excitations may drive 
strong jets perpendicular to any surrounding, large accretion disks.  
This process may have caused the strong jets visible near M87 and 
within the Seyfert 2 galaxies inNGC 5128 (Centaurus A), and Cygnus A 
clusters, as well as Hercules A.

At the same time, an extragalactic black hole acquisition within M87, 
and Centaurus, Cygnus, and Hercules A, would also have taken any 
objects, which previously had been quietly orbiting either black 
hole, and scattered them to the four winds. There is also the 
systematic difference in the velocities of the nearby companion star 
populations left over from the merger of two host galaxies containing 
the black holes, as well as the kinetic energy deposited by either 
black hole due to dynamical friction. All of these effects would 
result in a mass estimate exaggerated by a factor of thousands.  
It is possible that all such estimates for black holes, in the 
billions of solar masses, have been the result of companion black 
hole acquisition. The strong linear relation\cite{M98} between host 
galaxy mass and the misattributed, virial black hole mass follows 
trivially.

Thus there is no longer any problem of the time needed to form such 
massive objects, simply because they are not that massive, at best in 
the few tens of millions of solar masses, easily possible to build 
from globular cluster black holes of a few million solar masses each.
The recent\cite{aaa,ddd} ``image'' of the black hole in M87 is 
likely to be a large region between the two orbiting black holes 
($\sim$336 au diameter for 20 $\mu$arc s in M87) that has
been cleared of stars and luminous material by the periodic 
gravitational disturbance, rather than any event
horizon large enough to be resolved.

\section{Other supraluminal excitations, GRBs and redshifts}
\label{HTRO}

Many, if not all, transient events in the distant Universe, including 
outbursts of AGNs and QSOs, on timescales of days-to-months, 
generally accepted as black-hole driven, to fast radio bursts, on 
timescales of milliseconds, are likely a result of one type of 
supraluminal excitation or another, and violate the inverse square 
law in certain directions, relieving the need for the extreme amounts 
of associated energy. Gamma-ray bursts, which last up to 100 seconds, 
may result from supernova core collapse and a linearly accelerated 
supraluminal excitation (the vCc's on the path become increasingly 
open, intersecting with previous vCc's, and generating focused beams 
in a ring around the direction of motion).  

This occurs in a volume of circumstellar gas a dozen light-days 
distant ($d$), and is due to a beam collimated to about half a 
degree or so ($\delta\psi$), producing a delay \textit{range} of 
$d (\delta\psi)^2/2 \sim 50$ s.\cite{spot} But the 
\textit{afterglows} of gamma-ray bursts, many of which can last for 
hours, might consist, at least partially, and at some time when there 
is plasma at the right radius, pulsar emission.\cite{M12} So far, 
however, no pulsation has been detected in the afterglow of at least 
one GRB.\cite{jjb,RULLI}

Since about half of all Swift gamma-ray bursts are known to have 
afterglows, then at least half of GRB progenitors leave pulsars 
behind, and at most half are black-hole-producing events. Of those 
events which produce pulsars, the ones due to core-merger supernovae 
could possibly be used to derive the redshift, as 500 Hz may be a 
standard candle in initial spin frequency for such objects, because 
there is always too much (orbital) angular momentum in the merger 
process, so the spin rate is set by the branching of the Maclaurin 
and Jacoby solutions.\cite{StirlingColgate}

The optical signature of the neutron star-neutron star (NS-NS) merger 
in NGC 4993, GW170817, was still detectable four \textit{days} after 
the event, though several magnitudes fainter, more likely indicating 
an outburst of some sort (see, e.g., [\onlinecite{Fong17}] and 
references therein). It may be possible that the rapid rotation 
involved in NS-NS mergers, along with a remnant magnetic field, 
produce supraluminal excitations which may drive much of the observed 
outburst phenomena. \textit{Gravitational} supraluminal excitations 
of matter dense enough to, in turn, radiate strong focused 
gravitational waves, might be possible in NS-NS and even black 
hole-NS mergers.

\section{General Discussion}
\label{Pulsars}
\subsection{Pulsars}

We have presented evidence that at lesat half of all Parkes radio 
pulsars have been discovered because we are close to a favored 
orientation (nearly perpendicular to the rotation axes) where 
their radiated power diminishes only as the first power of distance, 
and have calculated the onset of this effect in one instance, 
confirming a long-predicted effect.\cite{Ar94,AR04}  

Observationally, the effect holds for these pulsars at distances up 
to 10 kpc.  Given the limited time resolution of almost all radio 
data, this can not be due to just the same energy concentrated into 
narrower and narrower pulses. Instead, out to some range of 
distances, more energy is concentrated into the same pulse width.

Interestingly, the strongest optical pulsations from SN 1987A,
observed during 1993, 6 Feb.\cite{Opacity} and 27 Aug. (at the 49.7 
kpc distance to the LMC), amounted\cite{M00,M00c} to a full 16\% 
(magnitude 20.5) of the m$_V$=18.5 central star which persisted from
about 1992-1995. Such pulsations correspond to an azimuth range of 
60$^{\circ}$ if this central source is entirely due to the pulsar, 
and more if only partly so. This implies, not coincidentally, that 
the pulsar was as bright as it \textit{could} get during the two 
observations, lending further support to the pulsar model discussed
in this work (but not for strongly-magnetized pulsars, such as the 
Crab).

\subsection{Supernovae}

The effects of focused beams, resulting from supraluminal 
excitations of polarization currents, are still generally 
unappreciated by the scientific community, but dominate SN 
disruption. Accounting for these in the early stages 
yields results that are highly 
anisotropic, but which are more limited in collimation than
the $\sim$100 to 1 visible for non-SN-producing mergers in
the direction of the Galactic center,\cite{Hey22} due to pulsar 
precession.\cite{M00,M00c} 
In SN 1987A these produced 0.62$\times$10$^{51}$ ergs 
(0.62 foe) from a 500-467.5 Hz-spin-rate drop.

In addition to the more obvious effects of anisotropy, we 
have from [\onlinecite{M12}], and considerations of this 
in Section \ref{SNdis}, a considerable fraction of the 
maximum apparent luminosity of SNe is due to polar jets 
propagating through (and at the same time, bunching up) 
polar-ejected material. This material is about a 
light-week and a half distant, though closer, and possibly 
brighter, for C-O stars due to the high atomic numbers in 
the jets and target material.  
This (time varying) contribution to the apparent luminosity 
depends on, and is variably delayed due to, the orientation 
angle of the pulsar rotational axis (more below).  

It has also already been more than a decade and a half since 
critical observations were made of local SNe Ia which were 
clearly below the width-luminosity relation of 
[\onlinecite{MMP}], established with serendipitously-discovered 
local Ia's, by nearly an order 
of magnitude,\cite{Benetti} \textit{with none lying above it
by the same factor}, 
and more than a dozen years since observers gave up on using 
them in that way.\cite{Siegfried} Correcting just a few such 
objects in a distant sample could easily wipe out the -28\% 
effect in luminosity, go on to overwhelm any reasonable 
correction for Malmquist bias (the last, best hope that less 
luminous distant SNe were due to a real effect), and still 
provide enough excess luminosity to imply a 
gravitationally-induced \textit{deceleration} in the Universe 
for $\Omega_M~\sim~ 0.99$. 

The reality is likely worse, with many more SNe Ia lying below 
the Width-Luminosity relation by such a large factor. Since we 
have shown that SNe are jet-dominated processes, in which those 
with jet/rotation axes at large angles to the line of sight 
have apparent luminosities an order of magnitude smaller 
than those with parallel axes (for SNe II the 75$^o$ angle for 
87A is consistent with its dim, $M_V \sim -16$), fainter SNe 
are overwhelmingly, numerically dominant (a few, otherwise 
apparently very luminous, Ia's are classified as Ic's when 
the bright, visible pole and diagnostic Si lines of the 
stellar remnant are obscured by the polar ejecta). 

In the end, this is a textbook
example of systematic biases in the measurement, in this case, 
of SN luminosity.  
It took almost a
decade to establish, in a more systematic way, a new local 
sample of Ia's to replace the initial,
and some of these were drastically dimmer than would 
be expected on the basis of the SN width-luminosity relation,
though Ia's with apparent luminosities another order of 
magnitude lower might have existed, but did not yet appear 
in this still sensitivity-limited, numerically small, new local 
sample. 



Thus it is clear that were SNe not so potentially luminous, 
they would be the last objects in the Universe anyone would
want to use as standard candles. But some \textit{are} that 
luminous, and fortunately, the very circumstellar material 
that complicates their use as standard candles may allow 
observation of many more relatively nearby SNe from before 
their instant of core collapse.\cite{M18,jg22} ``Bright 
Source 2'' in 87A 
was very likely\cite{M18} due to material ejected from a pole 
some months prior to core collapse, and was bright enough to 
detect in progenitors out to a few Mpc, and so these can 
serve as predictors of core collapse, as apparently 
happened in the case\cite{jg22} of SN 2020tlf.

A thorough study of a few relatively nearby SNe Ia just might 
lead to telltale details in these, that can also be measured 
in distant SNe, which can be used to determine the rotational 
orientation, as well as the physical properties of the
circumstellar material and the kinematics of the jets. So, 
in the end, after some very hard work, SNe Ia might actually 
be usable as standard candles.

But for now, the so-called ``Standard Model of Cosmology,'' 
with $\Omega_\mathrm{M} = 0.04,~\Omega_\mathrm{DM} = 0.24,$ 
$~\mathrm{and}~$ $\Omega_\mathrm{DE} = 0.72$, has no 
particular advantage over any other dogma, given its 
discrediting by \text{both} observation (20 years ago) and 
new understanding, and should be avoided also because 
of the baggage of ``dark energy'' and ``dark matter'' that 
comes with it. Indeed, there is no firm evidence for either, 
other than effects due to non-viriality and/or viriality 
attributed to the wrong mass, for the latter. There is also 
the problem of early star formation for low fractions of 
$\Omega_M$, which can be alleviated in ``bounce'' models with 
high $\Omega_M$, along with the problems of light elements
surviving the associated nucleosynthesis, 
and uniformity at the same time, without invoking ``Inflation'' 
(more baggage). The early formation of clusters, more important
than ever now with results from the James Webb Space Telescope,
and their high kinetics, may be explained by pulsar-driven 
jets.\cite{M12}

Still, students are being led down a primrose path of bad 
science (exploiting SNe for cosmology\cite{JAK} without first 
understanding\cite{des} them), ignorance (of the implications 
of SN 1987A as advanced in [\onlinecite{M12}] and
[\onlinecite{M09}] years before), and prejudice (some favoring 
a radical cosmology with a fervor that would shame the most 
zealous religious fanatic), by, now, a second or third 
generation of mentors, in part, perhaps, due to the 
persistence of supporting grants, which should be replaced by 
others which are much more generic in nature. 
    
Misunderstanding the nature of core-collapse SNe by everyone, 
particularly by those experts on the spectra and light curves 
of SNe, consulted to render judgement on an individual SN Ia 
which did not fit the ``pattern'' of anomalously dim Ia's 
to disqualify an 
offending candidate, is not the best way to produce objective 
science. This may also contribute to a distribution of 
luminosities of qualified SNe Ia not being consistent with 
\textit{any} known cosmology.\cite{RGV}

The issue of thermonuclear SNe Ia was raised a half 
century ago,\cite{Whelan} and persists even today. However, 
Wolf-Rayet stars with merging cores massing 1.4 M$_{\bigodot}$ 
or more, as happened within Sk -69$^{\circ}$202, a blue 
supergiant with plentiful hydrogen, will undergo 
core collapse and become SN Ia's themselves, thereby ruining 
the sample, no matter if thermonuclear SN Ia's, should they 
exist, can, by themselves, be used as standard candles.

\subsection{The Sun}

Closer to home, it may be possible that the 5-minute oscillations 
in the Sun\cite{Hill} also (sporadically) mimic a circularly 
supraluminal excitation. The circumference of the Sun is 14.7 
light-sec, so supraluminal excitations are possible for any 
harmonic number above 20 in these p-waves, which, in an 
already-magnetized body, have an electromagnetic consequence.  
Harmonics exist well above 60, which would imply $R$ values at 3 
or above. The more intermediate harmonics may put the initial 
focus close to, or just underneath the solar surface, possibly 
causing flares, and/or coronal mass ejections, and
possibly even\cite{Schaefer00} ``superflares.'' The 
highest harmonics can not dominate, or many of the flares and 
ejections would be polar. Whether the existence of such 
excitations will effect our estimates of $\textrm{T}_\textrm{C}$, 
its $\sim$15 million $^{\circ}$K core temperature, remains to be 
determined, (but if it does, the result may cast doubt on the 
existence of neutrino masses/flavor oscillations).

\subsection{Hercules X-1}

Parts of the twisted, tilted, accretion disk\cite{jap} around 
this 1.24 s X-ray pulsar will be illuminated by the pulsed, X-ray 
flux, thus producing supraluminal excitations, which may have 
resulted in very high energy pulsed radiation,\cite{D88,R88,L88,V89} 
observed in 1986, and upshifted in pulsed frequency by some 0.16\%, 
three times the maximum gain of the pulsar from Doppler shift.  Detailed discussion of this is too long to include here.

\subsection{Other effects}

The mechanism of focused radiation may also be important to 
gamma-ray bursts and their afterglows, AGN jets, neutron star 
mergers, and most, if not all, other transient events in the 
distant Universe. But to actually calculate these things to the 
end of the Universe, a greater range of precision in computation 
(256 bits) will be necessary. Until that time, no calculation of 
supernovae, or any other large object involving supraluminal 
excitations and their resulting focused beams will be possible. 
A few more orders of magnitude may be achieved for the supernova 
problem by iterating to find the minimum slope of the observer
time as an indicator, rather than fine-raster source times 
and histogramming the resulting observer times as plotted in 
Fig.~\ref{roundoga}, though this latter is visually more dramatic.

The search for effects of physics ``Beyond the Standard Model''
continues, but it is clear that the effects of supraluminal
excitations are: real (but were never considered until very
recently), responsible for many effects in the
distant Universe, and should themselves be considered as 
additions to the Standard Model of Physics.

\section{Conclusion}
\label{Conc}

Focused beams, produced by supraluminally undpated polarization 
currents driven by the neutron star/pulsar remnant, are the 
dominant disruption mechanism for most, if not all supernovae
(about a \textit{decade} down the road, this back-to-back, 
dual particle accelerator runs through all of the stellar 
material).  Nearly all progenitor stars are large enough so 
that the beam emergences cluster around the rotational poles, 
producing polar jets.

The jets will crash into material ejected previously by the 
same mechanism, acting through the increasing stellar core 
rotation rate, producing a significant contribution to the 
total SN luminosity which varies strongly with rotational 
orientation. The timing of this contribution also varies 
by several days, again according to the rotational 
orientation to the observer. They will not be useful as 
standard candles anytime soon, but there is some hope that
SNe can be understood in sufficient detail\cite{M18} to 
change this.  Until then, they provide no constraint on 
cosmological constants.  

Because the mechanism discussed here dominates the 
disruption of SNe, all previous calculations which 
successfully disrupted progenitors \textit{without} 
accounting for it are now embarrassingly incorrect, 
and will have to be modified to accommodate this new 
understanding of reality.

In addition, nearly all solitary progenitor stars spend 
their lives burning to heavier and heavier elements in 
their cores, and in consequence the moments of inertia of 
their magnetized cores will also fall and their rotation 
will speed up during their lifespans. Thus the same 
mechanism, that later disrupts SN progenitors, will cause 
material ejection in \textit{all} sufficiently large stars 
before core collapse (if that happens). It will also 
disrupt would-be ``Direct Collapse'' stars long before 
they achieve masses anywhere close to a billion solar.

Since there is no way to produce billion-solar-mass black
holes in the lifetime of the Universe, galaxies with jets 
and black holes thought to weigh in the billions of solar, 
such as M87, and Centaurus, Cygnus, and Hercules A, are 
likely the result of \textit{binary}, few-million solar 
mass black holes, the result of mergers and capture by 
dynamical friction (else, where \textit{are} the 
few-million-solar-mass black hole binaries, and how 
\textit{would} they appear?).



AGN jets likely do not originate directly from their
black holes, but rather from supraluminal excitations 
in the surrounding disk material, in turn produced by
magnetic fields moving supraluminally due to orbiting
magnetic bodies, and/or time varying gravitational 
fields from (most likely two) orbiting black holes 
with masses in the millions of solar.

In order to calculate these effects well to the ends of the
Universe, computers will have to be much more precise, 
possibly using as many as 256 bits for its constants and 
variables.

\section{Addendum}

Observations of the progenitor of SN 2020tlf reported in 
[\onlinecite{jg22}] showed an increase in luminosity for 
130 days prior to core-collapse, in good agreement with 
the ``five months'' estimate, derived from mass ejection 
due to supraluminal excitations, given in 
[\onlinecite{M18}].

In addition, the high energy neutrino sources from the 
Milky Way, observed by the IceCube detector,\cite{IC23}
might be completely accounted for by the young (rapidly
rotating) strongly magnetized pulsars close to the 
Galactic plane, acting through supraluminal exitations
in self-generated (pair-production) plasma, and the 
extragalactic flux\cite{IC13} might be attributed to 
the two such pulsars in the LMC, J0537-6910 and 
J0540-6919 (spinning at $\le \sim62$ Hz and $\le \sim20$ 
Hz), although more distant active galactic sources could
be far more luminous.\cite{FGH24} However, our 
galaxy has nothing like those sources.

Finally, the recent preoccupation with the ``Hubble
Tension,'' or the difference between the directly observed
value of $\rm{H_o}$ and that extrapolated from the Planck
spectrum using $\Lambda$CDM parameters,\cite{R21} seems 
pointless, no matter that it can be fine-tuned away
by the latest correction,\cite{MGJD22} given the unsuitability, 
established in this work, of using SNe Ia (or any other type of 
SN) as standard candles, and the resulting implied invalidity of 
$\Lambda$CDM. 

\begin{acknowledgments}

I would like to thank Eduardo B. Am\^ores of UEFS, Departamento 
de F\'isica Feira de Santana, CEP 44036-900, BA, Brazil, for
help with Cepheids and advice over the last several years.
I also gratefully acknowledge support for this work through the 
Los Alamos National Laboratory (LDRD) grants no. 20080085DR, 
``Construction and Use of Superluminal Emission Technology 
Demonstrators with Applications in Radar, Astrophysics, and 
Secure Communications,'' 20110320ER, ``Novel Broadband 
TeraHertz Sources, for Remote Sensing, Security, and
Spectroscopic Applications,'' and 20180352 ER, ''Scalable 
dielectric technology for VLF antennas.'' I would also like to 
thank John Singleton and Andrea Schmidt of Los Alamos National
Laboratory (LANL) for useful discussions, 
and Houshang \& Arzhang Ardavan, Pinaki Sengupta, Mario Perez, 
Todd Graves, \& Jesse Woodroffe (LANL) for their help. Finally 
I thank Hui Li of the Nuclear and Particle Physics, Astrophysics
and Cosmology group, also at LANL.

\end{acknowledgments}

\email{j.middleditch@gmail.com}


\begin{thebibliography}{15}
\bibitem{wind} With the exception of young, strongly magnetized 
pulsars such as the Crab, which can drive strong plasma winds 
within which other features of the pulse profile can be generated 
(see near the end of Section \ref{MI}).
\bibitem{scat} Interstellar scattering throws energy from 
this pulse into the neighboring $\pm$50 ms, which averages 
$\sim$20\% as high --- if this was \textit{not} included in 
the PKSMB S1400 flux estimate, then it would be closer to 
the most luminous.
\bibitem{W1400} At 1400 MHz -- if the spectrum of B1302 is 
steeper than most, then it will move up a bit in the order 
at the Westerbork 328 MHz frequency.
\bibitem{Narrow} However, narrower peaks are also easier to 
detect using reasonably sophisticated search methods.
\end{thebibliography}
\end{document}